\pdfoutput=1
\documentclass[12pt,fleqn]{article}

\usepackage{amsfonts,ulem,url} \usepackage{amssymb} \usepackage{amsmath}
\usepackage{amstext,verbatim,graphicx,ifthen,multirow,amsthm}

\usepackage{natbib}

\renewcommand{\thepage}{}
\renewcommand{\appendix}{\footnotesize\parindent 0cm\setcounter{equation}{0} 
\renewcommand{\theequation}{A.\arabic{equation}}
\setcounter{lemma}{0}\renewcommand{\thelemma}{A.\arabic{lemma}}}

\newcommand{\indep}{\perp\!\!\!\perp}

\newcommand{\calo}{{\cal O}}
\newcommand{\bw}{{\bf W}}
\newcommand{\bl}{{\bf L}}
\newcommand{\by}{{\bf Y}}
\newcommand{\bs}{{\bf S}}
\newcommand{\bu}{{\bf U}}
\newcommand{\bv}{{\bf V}}

\newcommand{\nn}{{\rm NN}}
\newcommand{\lasso}{{\rm LASSO}}
\newcommand{\rf}{{\rm RF}}

\newcommand{\ls}{{\rm ls}}

\newcommand{\mme}{\mathbb{E}}

\def\monthname{\ifcase\month\or
January\or February\or March\or April\or May\or June\or
July\or August\or September\or October\or November\or December\fi}

\renewcommand{\appendix}{\small\parindent 0cm\setcounter{equation}{0} 
\renewcommand{\theequation}{A.\arabic{equation}}
\setcounter{lemma}{0}\renewcommand{\thelemma}{A.\arabic{lemma}}
\setcounter{theorem}{0}\renewcommand{\thetheorem}{A.\arabic{theorem}}}

\begin{document}

\title{Machine Learning Methods Economists Should Know About\thanks{%
{\small We are grateful to Sylvia Klosin for comments.  This research was generously supported by ONR grant N00014-17-1-2131 and the Sloan Foundation.
}}}
\author{Susan Athey\thanks{{\small Professor of Economics, Graduate School of Business, Stanford University, SIEPR, and NBER, athey@stanford.edu. }} \and
 Guido W. Imbens\thanks{{\small Professor of Economics, Graduate School of Business and Department of Economics, Stanford University, SIEPR, and NBER, imbens@stanford.edu.}}
}
\date{
 \ifcase\month\or
January\or February\or March\or April\or May\or June\or
July\or August\or September\or October\or November\or December\fi \ \number%
\year
}
\maketitle

\begin{abstract}
We discuss the relevance of the recent Machine Learning (ML) literature for economics and econometrics. First we discuss the differences in goals, methods and settings between the ML  literature and the traditional econometrics and statistics literatures. Then we discuss some specific methods from the machine learning literature that we view as important for empirical researchers in economics. These include supervised learning methods for regression and classification, unsupervised learning methods, as well as matrix completion methods. Finally, we highlight newly developed methods at the intersection of ML and econometrics, methods that typically perform better than either off-the-shelf ML or more traditional econometric methods when applied to particular classes of problems, problems that include causal inference for average treatment effects, optimal policy estimation, and estimation of the counterfactual effect of price changes in consumer choice models.
\end{abstract}



\baselineskip=20pt\newpage \setcounter{page}{1}\renewcommand{\thepage}{[%
\arabic{page}]}\renewcommand{\theequation}{\arabic{section}.%
\arabic{equation}}

\section{Introduction}
\label{section:introduction}

In the abstract of his provocative 2001 paper in {\it Statistical Science} the Berkeley statistician Leo Breiman
writes about the difference between model-based versus algorithmic approaches to statistics:
\begin{quote} ``There are two cultures in the use of statistical modeling to reach conclusions from data. One assumes that the data are generated by a given stochastic data model. The other uses algorithmic models and treats the data mechanism as unknown.'' \citet{breiman2001statistical}, p199.\end{quote}
Breiman goes on to claim that: 
\begin{quote} ``The statistical community has been committed to the almost exclusive use of data models. This commitment has led to irrelevant theory, questionable conclusions, and has kept statisticians from working on a large range of interesting current problems. Algorithmic modeling, both in theory and practice, has developed rapidly in fields outside statistics. It can be used both on large complex data sets and as a more accurate and informative alternative to data modeling on smaller data sets. If our goal as a field is to use data to solve problems, then we need to move away from exclusive dependence on data models and adopt a more diverse set of tools.'' \citet{breiman2001statistical}, p199.\end{quote}
Breiman's characterization no longer applies to the field of statistics. The statistics community has by and large accepted the Machine Learning (ML) revolution that Breiman refers to as the algorithm modeling culture, and many textbooks discuss ML methods alongside more traditional statistical methods, {\it e.g.,} \citet{hastie2009elements} and \citet{efron2016computer}. Although the adoption of these methods in economics has been slower, they are now beginning to be widely used in empirical work, and are the topic of  a rapidly increasing methodological literature.
 In this paper we want to make the case that economists and econometricians also, as Breiman writes referring to the statistics community, ``need to move away from  exclusive dependence on data models and adopt a more diverse set of tools.'' We discuss some of the specific tools that empirical researchers would benefit from, and which we feel should be part of the standard graduate curriculum in econometrics if, as Breiman writes, and we agree with, ``our goal as a field is to use data to solve problems,'' if, in other words, we view econometrics as in essence, decision making under uncertainty ({\it e.g.,} \citet{chamberlain2000econometrics}), and if we wish to enable students to  be able to communicate effectively with researchers in other fields where these methods are routinely being adopted.
Although relevant more generally, the methods developed in the ML literature have been particularly successful in ``big data'' settings, where  we observe information on a large number of units, or many pieces of information on each unit, or both, and often outside the simple setting with a single cross-section of units.  For such settings, ML tools are becoming the standard across disciplines, and so the economist's toolkit needs to adapt accordingly, while preserving the traditional strengths of applied econometrics.
 
Why has the acceptance of ML methods been so much slower in economics compared to the broader statistics community? A large part of it may be the culture as Breiman refers to it. Economics journals emphasize the use of methods with formal properties of a type that many of the ML methods do not naturally deliver. This includes large sample properties of  estimators and tests, including  consistency, Normality, and efficiency. In contrast,  the focus in the machine learning literature is often on working properties of algorithms in specific settings, with the formal results of a different type, {e.g.,}  guarantees of error rates. There are typically fewer theoretical results of the type traditionally reported in econometrics papers, although recently there have been some major advances there (\citet{wager2017estimation, farrell2018deep}). 
There are no formal results that show that for supervised learning problems deep learning  or neural net methods are uniformly  superior to regression trees or random forests, and it appears unlikely that general results for such comparisons will soon be available, if ever. 

Although the ability to construct valid large sample confidence intervals is important in many cases, one should not out-of-hand dismiss methods that cannot not deliver them (or possibly, that  can not yet deliver them), if these methods have other advantages.
The demonstrated ability to outperform alternative methods on specific 
data sets  in terms of out-of-sample predictive power is valuable in practice, even though such performance is rarely explicitly acknowledged as  a goal, or assessed,  in econometrics.  As \citet{mullainathan2017machine} highlights, some substantive problems are naturally cast as prediction problems, and assessing their goodness of fit on a test set may be sufficient for the purposes of the analysis in such cases.  In other cases, the output of a prediction problem is an input to the primary analysis of interest, and statistical analysis of the prediction component beyond convergence rates is not needed.  On the other hand, there are also many settings where it is important to provide valid confidence intervals for a parameter of interest, such as an average treatment effect.  The degree of uncertainty captured by standard errors or confidence intervals may be a component in decisions about whether to implement the treatment.  We argue that in the future, as ML tools are more widely adopted, researchers should articulate clearly the goals of their analysis and why certain properties of algorithms and estimators may or may not be important.  

A major theme of this review is that even though there are cases where using simple of-the-shelf algorithms from the ML literature can be effective (see \citet{mullainathan2017machine} for a number of examples), there are also many cases where this is not the case. 
 Often the ML techniques require careful tuning and adaptation to effectively address the specific problems economists are interested in. Perhaps the most important type of adaptation is to exploit the structure of the problems, {\it e.g.,} the causal nature of many estimands, the endogeneity of variables, the configuration of data such as panel data, the nature of discrete choice among a set of substitutable products, or the presence of credible restrictions motivated by economic theory, such as monotonicity of demand in prices or other shape restrictions (\citet{matzkin1994restrictions, matzkin2007nonparametric}). 
 Statistics and econometrics have traditionally put much emphasis on these structures, and developed insights to exploit them, whereas ML has often put little emphasis on them.
 Exploiting these insights, both substantive and statistical, which, in a different form, is also seen in the careful tuning of ML techniques for specific problems such as image recognition, can greatly improve their performance.  Another type of adaptation involves changing the optimization criteria of machine learning algorithms to prioritize considerations from causal inference, such as controlling for confounders or discovering treatment effect heterogeneity.  Finally, techniques such as sample splitting (using different data to select models than to estimate parameters ({\it e.g.,} \citet{athey2016recursive, wager2017estimation}) and orthogonalization ({\it e.g.} \citet{chernozhukov2016double}) can be used to improve the performance of machine learning estimators, in some cases leading to desirable properties such as asymptotic normality of machine learning estimators ({\it e.g.} \citet{athey2017generalized, farrell2018deep}). 


In this paper, we discuss a list of tools that we feel should be be part of the empirical economists' toolkit and that we view should be covered in the core econometrics graduate courses.  Of course, this is a subjective list, and given the speed with which this literature is developing, the list will rapidly evolve. 
Moreover, we will not give a comprehensive discussion of these topics, rather we aim to provide an introduction to these methods that conveys the main ideas and insights, with references to more comprehensive treatments.
First on our list is 
nonparametric regression, or in the terminology of the ML literature, supervised learning for regression problems.
Second, 
supervised learning for classification problems, or closely  related, but not quite the same, nonparametric regression for discrete response models. This is the area where ML methods have perhaps had their biggest successes.
Third, unsupervised learning, or clustering analysis and density estimation. Fourth, we analyze estimates of heterogeneous treatment effects and optimal policies mapping from individuals' observed characteristics to treatments.  Fifth, we discuss ML approaches to  experimental design, where bandit approaches are starting to revolutionize effective experimentation especially in online settings.
Sixth, we discuss the matrix completion problem, including its application to causal panel data models and problems of consumer choice among a discrete set of products.  Finally, we discuss the analysis of text data.

We note that
there are a few other recent reviews of ML methods aimed at economists, often with more empirical examples and references to applications than we discuss here.  
\citet{varian2014big} is an early high level discussion of a selection of important ML methods.
\citet{mullainathan2017machine} focus on the benefits of  supervised learning methods for regression, and discuss the prevalence of problems in economics where prediction methods are appropriate. 
\citet{athey2017beyond} and \citet{athey2017impact} provides a broader perspective with more emphasis on recent developments in adapting ML methods for causal questions and general implications for economics.
\citet{gentzkow2017text} provide an excellent recent discussion of methods for text analyses with a focus on economics applications.
In the computer science and statistics literatures there are also a number of excellent textbooks, with different levels of accessibility to researchers with a social science background, including \citet{ efron2016computer},  \citet{hastie2009elements}, which is a more comprehensive text from a statistics perspective, and
\citet{burkov100} which is a very accessible introduction,
 \citet{alpaydin2009introduction}, and \citet{knox2018machine}, which all take more of a computer science perspective.

\section{Econometrics and Machine Learning: Goals,  Methods, and Settings}

In this section we introduce some of the general themes of this paper. What are the differences in the goals and concerns of traditional econometrics and the machine learning literature, and how do these goals and concerns affect the choices between specific methods?

\subsection{Goals}

The traditional approach in econometrics, as exemplified in leading texts such as
\citet{wooldridge2010econometric,  angrist2008mostly, greene2000econometric} is to specify a target, an estimand, that is a functional of a joint distribution of the data. 
Often the target is a parameter of a statistical model that describes the distribution of a set of variables (typically conditional on some other variables) in terms of a set of parameters, which can be a finite or infinite set. Given a random sample from the population of interest the parameter of interest and the nuisance parameters are estimated by finding the parameter values that best fit the full sample, using an objective function such as the sum of squared errors, or the likelihood function. The focus is on the quality  of the  estimators of the target, traditionally measured through large sample efficiency. Often there is also interest in constructing confidence intervals. Researchers typically report point estimates and standard errors.

In contrast, in the ML literature the focus is typically on developing algorithms (a widely cited paper, 
\citet{wu2008top}, has the title ``Top 10 algorithms in data mining''). The goal for the algorithms is typically to make predictions about some variables given others, or classify units on the basis of limited information, for example to classify handwritten digits on the basis of pixel values. 

In a very simple example, 
suppose we model the conditional distribution of some  outcome $Y_i$ given a vector-valued regressor or feature $X_i$. Suppose we are confident that
\[ Y_i|X_i\sim{\cal N}(\alpha+\beta^\top X_i,\sigma^2).\]
We could estimate $\theta=(\alpha,\beta)$ by least squares, that is, as
\[ (\hat\alpha_\ls,\hat\beta_\ls)=\arg\min_{\alpha,\beta}
\sum_{i=1}^N \left( Y_i-\alpha-\beta^\top X_i\right)^2.\]
Most introductory econometrics texts would focus on the least squares estimator  without much discussion. 
If the model is correct, the least squares estimator has well known attractive properties: it is unbiased, it is the best linear unbiased estimator, it is the maximum likelihood estimator, and so has large sample efficiency properties. 

In ML settings the goal may be to make a prediction for the outcome for a new units on the basis of their regressor values.
 Suppose we are interested in predicting
the value of $Y_{N+1}$ for a new unit $N+1$, on the basis of the regressor values for this new unit, $X_{N+1}$.
Suppose we restrict ourselves to  linear predictors, so that the prediction is
\[ \hat Y_{N+1}=\hat\alpha+\hat\beta^\top X_{N+1},\]
for some estimator $(\hat\alpha,\hat\beta)$.
The loss associated with this decision may be the squared error
\[ \left (Y_{N+1}-\hat Y_{N+1}\right)^2.\]
The question now is to come up with  estimators $(\hat\alpha,\hat\beta)$ that have good properties associated with this loss function.  This need not be the least squares estimator. In fact, when the dimension of the features exceeds two, we know from decision theory that we can do better in terms of expected squared error than the least squares estimator. The latter is not admissible, that is, there are other estimators that dominate the least squares estimator.

\subsection{Terminology}

One source of confusion is the use of new terminology in the ML for concepts that have well-established labels in the older literatures. In the context of a regression model the sample used to estimate the parameters is often referred to as the {\it training} sample. Instead of estimating the model, it is being {\it trained}. Regressors, covariates, or predictors are referred to as {\it features}. Regression parameters are sometimes referred to as {\it weights}. Prediction problems are divided into {\it supervised learning problems} where we observe both the predictors/features $X_i$ and the outcome $Y_i$, and {\it unsupervised learning problems} where we only observe the $X_i$ and try to group them into clusters or otherwise estimate their joint distribution. Unordered discrete response problems are generally referred to as {\it classification problems}.

\subsection{Validation and Cross-validation}
\label{section:validation}

In most discussions on linear regression in econometric textbooks there is little emphasis on model validation. The form of the regression model, be it parametric or nonparametric, and the set of regressors, is assumed to be given from the outside, {\it e.g.,}  economic theory. Given this specification, the task of the researcher is to estimate the unknown parameters of this model. Much emphasis is on doing this estimation step efficiently,  typically operationalized through definitions of large sample efficiency. If there is discussion of model selection, it is often in the form of testing null hypotheses concerning the validity of a particular model, with the implication that there is a true model that should be selected and used for subsequent tasks.

Consider the regression example in the previous subsection. Let us assume that we are interested in predicting the outcome for a new unit, randomly drawn from the same population as our sample was drawn from. As an alternative to estimating the linear model with an intercept, and a scalar $X_i$, we could estimate the model with only an intercept. Certainly if $\beta=0$, that model would lead to better predictions. By the same argument, if the true value of $\beta$ were close, but not exactly equal, to, zero, we would still do better leaving $X_i$ out of the regression.  Out-of-sample cross-validation can help guide such decisions.
  There are two components of the problem that are important for this ability. First, the goal is predictive power, rather than estimation of a particular structural or causal parameter. Second, the method uses out-of-sample comparisons, rather than in-sample goodness-of-fit measures. This ensures that we obtain unbiased comparisons of the fit.

\subsection{Over-fitting, Regularization, and Tuning Parameters}
\label{section:regularization}

The ML literature is much more concerned with over-fitting than the standard statistics or econometrics literature. Researchers attempt to select flexible models that fit well, but not so well that out-of-sample prediction is compromised. There is much less emphasis on formal results that particular methods are superior in large samples (asymptotically), instead methods are compared on specific data sets to see ``what works well.'' A key concept is that of {\it regularization}.
As Vapnik writes,
\begin{quote} ``Regularization theory was one of the first signs of the existence of intelligent inference''
(\citet{vapnik1998statistical}, p.)\end{quote}
Consider a setting with a large set of models that differ in their complexity, measured for example as  the number of unknown parameters in the model, or, more subtly, through the
 the Vapnik--Chervonenkis (VC) dimension that measures  the capacity  or complexity of a space of models.
Instead of directly optimizing an objective function, say minimizing the sum of squared residuals in a least squares regression setting, or maximizing the logarithm of the likelihood function, a  term is added to the objective function to penalize the complexity of the model. There are antecedents of this practice in the traditional econometrics and statistics literature. One is that in likelihood settings researchers sometimes add a term to the logarithm of the likelihood function equal to minus the logarithm of the sample size times the number of free parameters divided by two, leading to the {\it Bayesian Information Criterion}, or simply the number of free parameters, the {\it Akaike Information Criterion}. In Bayesian analyses of regression models the use of a prior distribution on the regression parameters, centered at zero, independent accross parameters with a constant prior variance, is another way of regularizing estimation that has a long tradition.
The difference with the modern approaches to regularization is that they are  more data driven, with the amount of regularization determined explicitly by the out-of-sample predictive performance rather than by, for example, a subjectively chosen prior distribution.

Consider a linear regression model with $K$ regressors,
\[ Y_i|X_i\sim{\cal N}\left(\beta^\top X_{i},\sigma^2\right).\]
Suppose we also have a prior distribution for the  the slope coefficients $\beta_k$, with the prior for $\beta_k$, $ {\cal N}(0,\tau^2)$, and independent of $\beta_{k'}$ for any $k\neq k'$. (This may be more plausible if we first normalize the features and outcome to have mean zero and unit variance. We assume this has been done.) Given the value for the variance of the prior distribution, $\tau^2$, the posterior mean for $\beta$ is the solution to
\[ \arg\min_{\beta}\sum_{i=1}^N \left(Y_i-\beta^\top X_{i}\right)^2
+\frac{\sigma^2}{\tau^2} \|\beta\|_2^2,\]
where $\|\beta\|_2=\left(\sum_{k=1}^K \beta_k^2\right)^{1/2}$.
One version of an ML approach to this problem is to estimate $\beta$ by minimizing
\[ \arg\min_{\beta}\sum_{i=1}^N \left(Y_i-\beta^\top X_{i}\right)^2
+\lambda \|\beta\|_2^2.\]
The only difference is in the way the penalty parameter $\lambda$ is chosen. In a formal Bayesian approach this reflects the (subjective) prior distribution on the parameters, and it would be chosen {\it a priori}. In an ML approach $\lambda$ would be chosen through out-of-sample cross-validation to optimize the out-of-sample predictive performance. This is closer to an Empirical Bayes approach where the data are used to estimate the prior distribution ({\it e.g.}, \citet{morris1983parametric}).

\subsection{Sparsity}

In many settings in the ML literature the number of features is substantial, both in absolute terms and relative to the number of units in the sample. However, there is often a sense that many of the features are of minor importance, if not completely irrelevant. The problem is that we may not know {\it ex ante} which of the features matter, and which can be dropped from the analysis without  substantially  hurting the predictive power.

\citet{hastie2009elements, hastie2015statistical} discuss what they call the {\it sparsity principle}:
\begin{quote}``Assume that the underlying true signal is sparse and we use an $\ell_1$ penalty to try to recover it. If our assumption is correct, we can do a good job in recovering the true signal.  ... But if we are wrong---the underlying truth is not sparse in the chosen bases---then the $\ell_1$ penalty will not work well. However, in that instance, no method can do well, relative to the Bayes error.'' (\citet{hastie2015statistical}, page 24).\end{quote}
Exact sparsity is in fact stronger than is necessary, in many cases it is sufficient to have approximate sparsity where most of the explanatory variables have very limited explanatory power, even if not zero, and only a few of the features  are of substantial importance (see, for example, \citet{belloni2014jep}).

Traditionally in the empirical literature in social sciences researchers limited the number of explanatory variables by hand, rather than choosing them in a data-dependent manner. Allowing the data to play a bigger role in the variable selection process appears a clear improvement, even if the assumption that the underlying process is at least approximately sparse is still a very strong one, and even if inference in the presence of data-dependent model selection can be challenging.

\subsection{Computational Issues and Scalability}\label{section:scale}

Compared to the traditional statistics and econometrics literatures the ML literature is much more concerned with computational issues and the ability to implement estimation methods with large data sets. Solutions that may have attractive theoretical properties in terms of statistical efficiency but that do not scale well to large data sets are often discarded in favor of methods that can be implemented easily in very large data sets. This can be seen in the discussion of the relative merits of LASSO versus subset selection in linear regression settings. In a setting with a large number of features that might be included in the analysis, subset selection methods focus on selecting a subset of the regressors and then estimate the parameters of the regression function by least squares. 
However, LASSO has computational advantages. It can be implemented by adding a penalty term that is proportional to the sum of the absolute values of the parameters. A major attraction of LASSO is that there are effective methods for calculating the LASSO estimates with the number of regressors in the millions. Best subset selection regression, on the other hand, is an NP-hard problem. Until recently it was thought that this was only feasible in settings with the number of regressors in the 30s, although current research
(\citet{bertsimas2016best}) suggests it may be feasible with the number of regressors in the 1000s.
This has reopened a new, still unresolved, debate on the relative merits of LASSO versus best subset selection (see
\citet{hastie2017extended}) in settings where both are feasible. There are some indications that in settings with a low signal to noise ratio, as is common in many social science applications, LASSO may have better performance, although there remain many open questions. In many social science applications the scale of the problems is such that best subset selection is also feasible, and the computational issues may be less important than these substantive aspects of the problems.

A key computational optimization  tool used in many ML methods is Stochastic Gradient Descent (SGD,  \citet{friedman2002stochastic, bottou1998online, bottou2012stochastic}).   It is used in a wide variety of settings, including in optimizing neural networks and estimating models with many latent variables (e.g., \citet{ruiz2017shopper}). The idea is very simple.  Suppose that the goal is to estimate a parameter
$\theta$, and the estimation approach entails finding the value $\hat\theta$ that minimizes an empirical loss function, where $Q_i(\theta)$ is the loss for observation $i$, and the overall loss is the sum $\sum_i Q_i(\hat\theta_k)$, with derivative $ \sum_i\nabla Q_i(\hat\theta)$.
Classic gradient decscent  methods involve an iterative approach, where $\hat\theta_k$ is updated from $\hat\theta_{k-1}$ as follows:
\[\theta_k = \theta_{k-1} - \eta_k \frac{1}{N} \sum_i  \nabla Q_i(\hat\theta),\]
where $\eta_k$ is the learning rate, often chosen optimally through line search.  More sophisticated optimization methods multiply the first derivative with the inverse of the matrix of second derivatives or estimates thereof.

The challenge with this approach is that it can be computationally expensive. The computational cost is in evaluating the full derivative $\sum_i\nabla Q_i$, and even more in optimizing the learning rate $\eta_k$.
The idea behind SGD is that  it is better to take many small steps that are noisy but on average in the right direction, than it is to spend equivalent computational cost in very accurately figuring out in what direction to take a single small step.
More specifically, SGD, uses the fact that the average of $\nabla Q_i$ for a random subset of the sample  is an unbiased (but noisy) estimate of the  gradient.   For example, dividing the data randomly into ten subsets or batches, with $B_i\in\{1,10\}$ denoting the subset unit $i$ belongs to, one could do ten steps of the type
\[\theta_k = \theta_{k-1} - \eta_k  \frac{1}{N/10} \sum_{i:B_i=k}\nabla  Q_i(\hat\theta_k),\]
with a deterministic learning rate $\eta_k$.
After the ten iteractions one could reshuffle the dataset and then repeat.  If the learning rate $\eta_k$  decreases at an appropriate rate, under relatively mild assumptions, SGD converges almost surely to a global minimum when the objective function is convex or pseudoconvex, and otherwise converges almost surely to a local minimum.  See \citet{bottou2012stochastic} for an overview and practical tips for implementation.

The idea can be pushed even further in the case where $\nabla Q_i(\theta)$ is itself an expectation.  We can consider evaluating $\nabla Q_i$ using Monte Carlo integration.  But, rather than taking many Monte Carlo draws to get an accurate approximation to the integral, we can instead take a small number of draws, or even a single draw. 
This type of approximation is used in economic applications in \citet{ruiz2017shopper} and \citet{Hartford2016}.

\subsection{Ensemble Methods and Model Averaging}

Another key feature of the machine learning literature is the use of model averaging and ensemble methods ({\it e.g.,} \citet{dietterich2000ensemble}). In many cases a single model or algorithm does not perform as well as a combination of possibly quite different models, averaged using weights (sometimes called {\it votes}) obtained by optimizing out-of-sample performance. 
A striking example is the Netflix Prize Competition (\citet{bennett2007netflix}), where all the top contenders use combinations of  models, often averages of many models (\citet{bell2007lessons}).
There are two related ideas in the traditional econometrics literature. Obviously Bayesian analysis implicitly averages over the posterior distribution of the parameters. Mixture models are also used to combine different parameter values in a single prediction. However, in both cases this model averaging involves averaging over similar models, typically with the same specification, and only different in terms of parameter values. In the modern literature, and in the top entries in the Netflix competition, the models that are averaged over can be quite different, and the weights are  obtained by optimizing out-of-sample predictive power, rather than in-sample fit.

For example, one may have three  predictive models, one  based on a random forest, leading to predictions $\hat Y_i^{\rm RF}$, one based on a neural net, with predictions $\hat Y^{\rm NN}_i$, and one based on a linear model estimated by LASSO, leading to $\hat Y_i^{\rm LASSO}$. Then, using a test sample, one can choose weights $p^\rf$, $p^\nn$, and $p^\lasso$, by minimizing the sum of squared residuals in the test sample:
\[ (\hat p^\rf,\hat p^\nn,\hat p^\lasso)=\arg\min_{p^\rf,p^\nn,p^\lasso}\sum_{i=1}^{N^{\rm test}}
\left( Y_i-p^\rf \hat Y_i^\rf-p^\nn \hat Y_i^\nn-p^\lasso \hat Y_i^\lasso\right)^2, \]
\[{\rm subject\ to\ }p^\rf+p^\nn+p^\lasso=1, \ \ {\rm and}\ p^\rf,p^\nn,p^\lasso\geq 0.\]
One may also estimate weights based on regression of the outcomes in the test sample on the predictors from the different models without imposing that the weights sum to one and are non-negative.
Because random forests, neural nets, and lasso have distinct strengths and weaknesses, in terms of how well they deal with the presence of irrelevant features, nonlinearities, and interactions. As a result  averaging over these models may lead to out-of-sample predictions that are strictly better than predictions based on a single model.

In a panel data context \citet{athey2019ensemble} use ensemble methods combining various forms of 
synthetic control and  matrix completion methods and find that the combinations outperform the individual methods.

\subsection{Inference}

The ML literature has focused heavily on out-of-sample performance as the criterion of interest. This has come at the expense of one of the concerns that the statistics and econometrics literature have traditionally focused on, namely
 the ability to do inference, {\it e.g.,} construct confidence intervals that are valid, at least in large samples. Efron and Hastie write:
\begin{quote} ``Prediction, perhaps because of its model-free nature, is an area where algorithmic developments have run far ahead of their inferential justification.'' (\citet{efron2016computer}, p. 209)\end{quote}
Although there has recently been substantial progress in the development of methods for inference for low-dimensional functionals in specific settings ({\it e.g.}, \citet{wager2017estimation} in the context of random forests, and \citet{ farrell2018deep} in the context of neural networks), it remains the case that for many methods it is currently impossible to construct confidence intervals that are valid, even if only asymptotically.  A question is whether this ability to construct confidence intervals is as important as the traditional emphasis on it in the econometric literature suggests. For many decision problems it may be that prediction is of primary importance, and inference is at best of secondary  importance.
Even in cases where it is possible to do inference, it is important to keep in mind that the requirements that ensure this ability  often come at the expense of predictive performance. One can see this tradeoff  in traditional kernel regression, where the  bandwidth that optimizes expected squared error balances the tradeoff between the square of the bias and the variance, so that the optimal estimators have an asymptotic bias that invalidates the use of standard confidence intervals. This can be fixed by using a bandwidth that is smaller than the optimal one, so that the asymptotic bias vanishes, but it does so explicitly at the expense of increasing the variance.

\section{Supervised Learning for Regression Problems}
\label{section:supervised}

One of the canonical problems in both the ML and econometric literatures is that of 
estimating the conditional mean of  a scalar outcome given a set of of covariates or features. Let $Y_i$ denote the outcome for unit $i$, and let $X_i$ denote the $K$-component vector of covariates or features. The conditional expectation is
\[ g(x)=\mme[Y_i|X_i=x].\]
 Compared to the traditional econometric textbooks ({\it e.g.,} \citet{angrist2008mostly, greene2000econometric, wooldridge2010econometric}) there are some conceptual differences with the ML literature (see the discussion in \citet{mullainathan2017machine}). In the settings considered in the ML literature there are often many covariates, sometimes  more than there are units in the sample. There is no presumption in the ML literature that the conditional distribution of the outcomes given the covariates follows a particular parametric model. The derivatives of the conditional expectation for each of the covariates, which in the linear regression model correspond to the parameters, are not of intrinsic interest. Instead the focus is on out-of-sample predictions and their accuracy. Furthermore, there is less of a sense that the conditional expectation is monotone in each of the covariates compared to many economic applications. Often there is concern that the  conditional expectation may be an extremely non-monotone function with some higher order interactions of substantial importance.

The econometric literature on estimating the conditional expectation is also huge.
Parametric methods for estimating $g(\cdot)$ often used least squares. Since the work by \citet{bierens1987kernel},  kernel regression methods have become a popular alternative when more flexibility is required, with subsequently series or  sieve methods gaining interest (see \citet{chen2007large} for a survey). These methods have well established large sample properties, allowing for the construction of confidence intervals. Simple non-negative kernel methods are viewed as performing very poorly  in settings with high-dimensional covariates, with the difference $\hat g(x)-g(x)$ of order $O_p(N^{-1/K})$. This rate can be improved by using higher order kernels and assuming the existence of many derivatives of $g(\cdot)$, but practical experience with high-dimensional covariates has not been satisfactory for these methods, and applications of kernel methods in econometrics are generally limited to low-dimensional settings.

The differences in performance between some of the traditional methods such as kernel regression and the modern methods such as random forests are particularly pronounced in sparse settings with a large number of more or less irrelevant covariates. Random forests are effective at picking up on the sparsity and ignoring the irrelevant features, even if there are many of them, while the traditional implementations of kernel methods essentially waste degrees of freedom on accounting for these covariates. Although it may be possible to adapt kernel methods for the presence of irrelevant covariates by allowing for covariate specific bandwidths, in practice there has been little effort in this direction. A second issue is that the modern methods are particularly good at detecting severe nonlinearities and high-order interactions. The presence of such high-order interactions in some of the success stories of these methods should not blind us to the fact that with many economic data we expect high-order interactions to be of limited importance. If we try to predicting earnings for individuals, we expect the regression function to be monotone in many of the important predictors such as education and prior earnings variables, even for homogenous subgroups. This means that models based on linearizations may do well in such cases relative to other methods, compared to settings where monotonicity is fundamentally less plausible, as, for example, in an image recognition problem. This is also a reason for the superior performance of locally linear random forests (\citet{friedberg2018local}) relative to standard random forests.

We discuss four specific sets of methods, although there are many more, including variations on the basic methods. First, we discuss methods where the class of models considered is linear in the covariates, and the question is solely about regularization. Next we discuss methods based on partitioning the covariate space using regression trees and random forests. In the third subsection we discuss neural nets, which were the focus on of a small econometrics literature in the 1990s (\citet{white1992artificial,  hornik1989multilayer}), but more recently has become a very prominent literature in ML in various subtle reincarnations. Then we discuss boosting as a general principle. 

\subsection{Regularized Linear Regression: Lasso, Ridge, and Elastic Nets}

Suppose we consider approximations to the conditional expectation that have a linear form
\[ g(x)=\beta^\top x=\sum_{k=1}^K \beta_kx_k,\]
after the covariates and the outcome are demeaned, and the covariates are normalized to have unit variance.
The traditional method for estimating the regression function in this case is least squares, with
\[ \hat\beta^\ls=\arg\min_{\beta}
\sum_{i=1}^N \left( Y_i-\beta^\top X_i\right)^2.\]
However, if the number of  covariates $K$ is large relative to the number of observations $N$ the least squares estimator $\hat\beta^\ls_k$ does not even have particularly good repeated sampling properties as an estimator for $\beta_k$, let alone good predictive properties. In fact, with $K\geq 3$ the least squares estimator is not even admissible and is dominated by estimators that shrink towards zero. With $K$ very large, possibly even exceeding the sample size $N$, the least squares estimator has particularly poor properties, even if the conditional mean of the outcome given the covariates is in fact linear.

Even with $K$ modest in magnitude, the predictive properties of the least squares estimator may be inferior to those of estimators that use some amount of regularization. 
One common form of regularization is to add a penalty term that shrinks the $\beta_k$ towards zero, and minimize
\[ \arg\min_{\beta}
\sum_{i=1}^N \left( Y_i-\beta^\top X_i\right)^2+
\lambda \left( \|\beta\|_q\right)^{1/q}.\]
where $\|\beta\|_q=\sum_{k=1}^K|\beta_k|^q$.
For $q=1$ this corresponds to LASSO (\citet{tibshirani1996regression}).
For $q=2$ this corresponds to ridge regression (\citet{hoerl1970ridge}). As $q\rightarrow 0$, the solution penalizes the number of non-zero covariates, leading to best subset regression (\citet{miller2002subset, bertsimas2016best}).
In addition there are many hybrid methods and modifications, including elastic nets which combines penalty terms from LASSO and ridge (\citet{zou2005regularization}), the relaxed lasso, which combines least squares estimates from the subset selected by LASSO and the LASSO estimates themselves 
(\citet{meinshausen2007relaxed}), Least Angle Regression (\citet{efron2004least}), the Dantzig Selector  (\citet{candes2007dantzig}),  the Non-negative Garrotte (\citet{breiman1993better}) and  others.

There are a couple of important conceptual differences between these three special cases, subset selection, LASSO, and ridge regression. 
See for a recent discussion \citet{hastie2017extended}.
Both best subset and LASSO lead to solutions with a number of the regression coefficients exactly equal to zero, a {\it sparse} solution. For the ridge estimator on the other hand all the estimated regression coefficients will generally differ from zero. It is not always important to have a sparse solution, and often the variable selection that is implicit in these solutions is over-interpreted. 
Second, best subset regression is computationally hard (NP-hard), and as a result not feasible in settings with $N$ and $K$ large, although recently  progress has been made in this regard (\citet{ bertsimas2016best}).
LASSO and ridge have a Bayesian interpretation.  Ridge regression gives the posterior mean and mode under a Normal model for the conditional distribution of $Y_i$ given $X_i$, and Normal prior distributions for the parameters. LASSO gives the posterior mode given Laplace prior distributions.
However,
in contrast to formal Bayesian approaches, the coefficient $\lambda$ on the penalty term  is in the modern literature choosen through out-of-sample crossvalidation rather than subjectively through the choice of prior distribution. 


\subsection{Regression Trees and  Forests}

Regression trees (\citet{breiman1984classification}), and their extension random forests (\citet{Breiman2001}) have become very popular and effective methods for flexibly estimating regression functions in settings where out-of-sample predictive power is important. They are considered to have great out-of-the-box performance without requiring subtle tuning. Given a sample $(X_{i1},\ldots,X_{iK},Y_i)$, for $i=1,\ldots,N$, the idea is to split the sample into subsamples,
and estimate the regression function within the subsamples simply as the average outcome. The splits are sequential and based on a single covariate $X_{ik}$ at a time exceeding a threshold $c$. 
Starting with the full training sample, consider a split based on feature or covariate $k$, and threshold  $c$. The sum of  in-sample squared errors before the split was
\[ Q=\sum_{i=1}^N \left (Y_i-\overline{Y}\right)^2,\hskip1cm {\rm where} \ \overline{Y}=\frac{1}{N}\sum_{i=1}^N Y_i.\]
After a split based on covariate $k$ and threshold $c$  the sum of in-sample squared errors is
\[ Q(k,c)=\sum_{i:X_{ik}\leq c} \left(Y_i-\overline{Y}_{k,c,l}\right)^2+
\sum_{i:X_{ik}> c} \left(Y_i-\overline{Y}_{k,c,r}\right)^2,\]
where (with ``l'' and ``r'' denoting ``left'' and ``right''),
\[ \overline{Y}_{k,c,l}=\sum_{i:X_{ik}\leq c} Y_i\Bigl/\sum_{i:X_{ik}\leq c} 1,
\hskip1cm {\rm and}\ \ 
 \overline{Y}_{k,c,r}=\sum_{i:X_{ik}> c} Y_i\Bigl/\sum_{i:X_{ik}> c} 1,\]
 are the average outcomes in the two subsamples.
We split the sample using the covariate $k$ and threshold $c$ that minimize the average squared error $Q(k,c)$ over all covariates $k=1,\ldots,K$ and all thresholds $c\in(-\infty,\infty)$. We then repeat this, now optimizing also over the subsamples or leaves. At each split the average squared error is further reduced (or stays the same).  We therefore need some regularization to avoid the overfitting that would result from splitting the sample too many times. One approach is to add a penalty term to the sum of squared residuals that is linear in the number of subsamples (the {\it leaves}). The coefficient on this penalty term is then chosen through cross-validation.  In practice, a very deep tree is estimated, and then {\it pruned} to a more shallow tree using cross-validation to select the optimal tree depth. The sequence of first growing followed by pruning the tree avoids splits that may be missed because  their benefits rely on subtle interactions.

An advantage of a single tree is that it is  easy to explain and interpret results.  Once the tree structure is defined, then the prediction in each leaf is a sample average, and the standard error of that sample average is easy to compute.  However, it is not in general true that the sample average of the mean within a leaf is an unbiased estimate of what the mean would be within that same leaf in a new test set.  Since the leaves were selected using the data, the leaf sample means in the training data will tend to be more extreme (in the sense of being different from the overall sample mean) than in an independent test set.  \citet{athey2016recursive} suggest sample splitting as a way to avoid this issue.  If a confidence interval for the prediction is desired, then the analyst can simply split the data in half.  One half of the data is used to construct a regression tree.  Then, the partition implied by this tree is taken to the other half of the data where the sample mean within a given leaf is an unbiased estimate of the true mean value for the leaf. 

Although trees are easy to interpret, it is important not to go too far in interpreting the structure of the tree, including the selection of variables used for the splits.  Standard intuitions from econometrics about ``omitted variable bias'' can be useful here. 
Particular covariates that have strong associations with the outcome may not show up in splits because the tree splits on covariates highly correlated with those covariates.

\begin{center}
\begin{tabular}{ccc}
\includegraphics[width = 0.4\textwidth, trim=15mm 40mm 20mm 40mm, clip = TRUE]{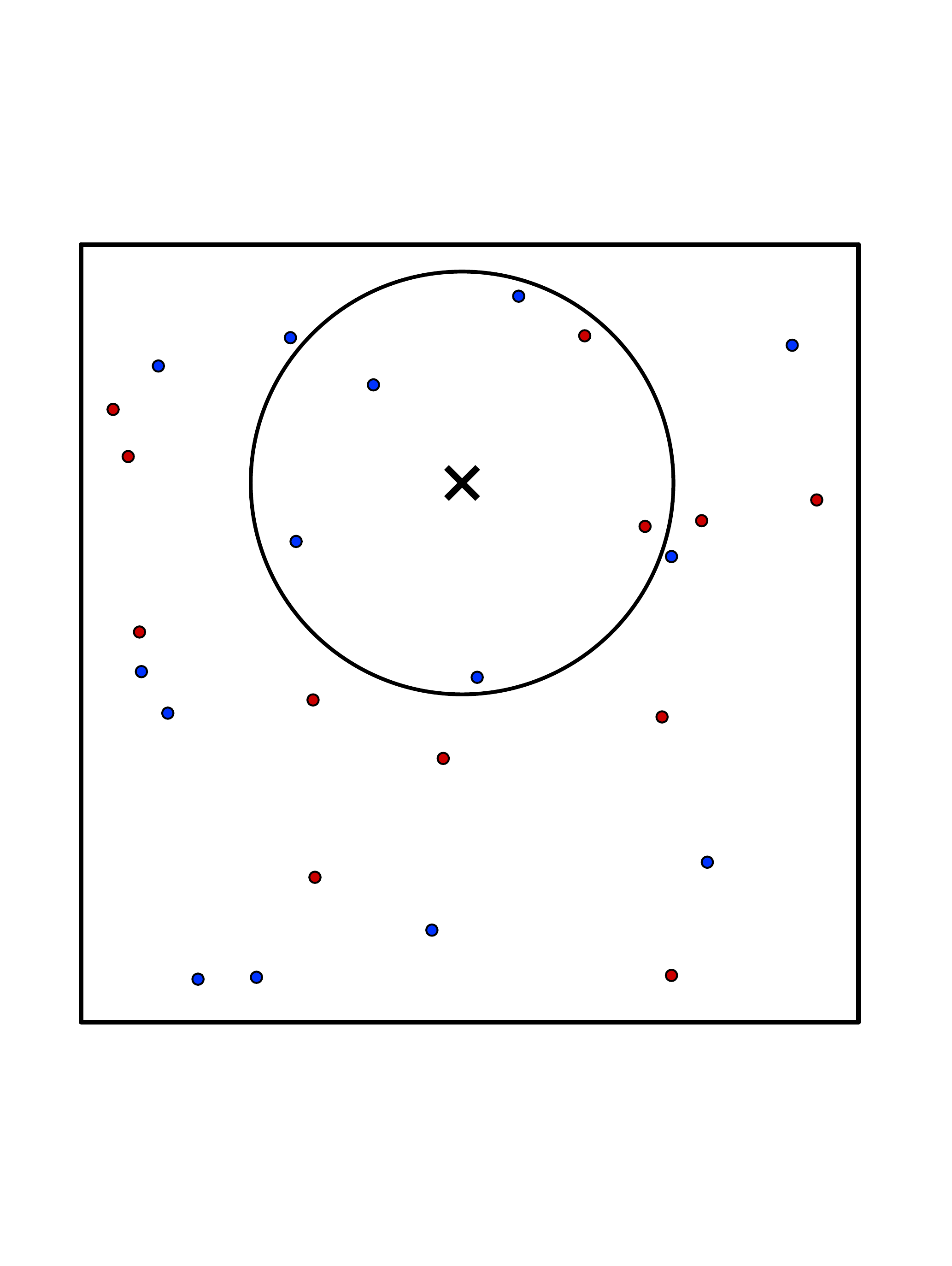} &
\hspace{0.05\textwidth} &
\includegraphics[width = 0.4\textwidth, trim=15mm 40mm 20mm 40mm, clip = TRUE]{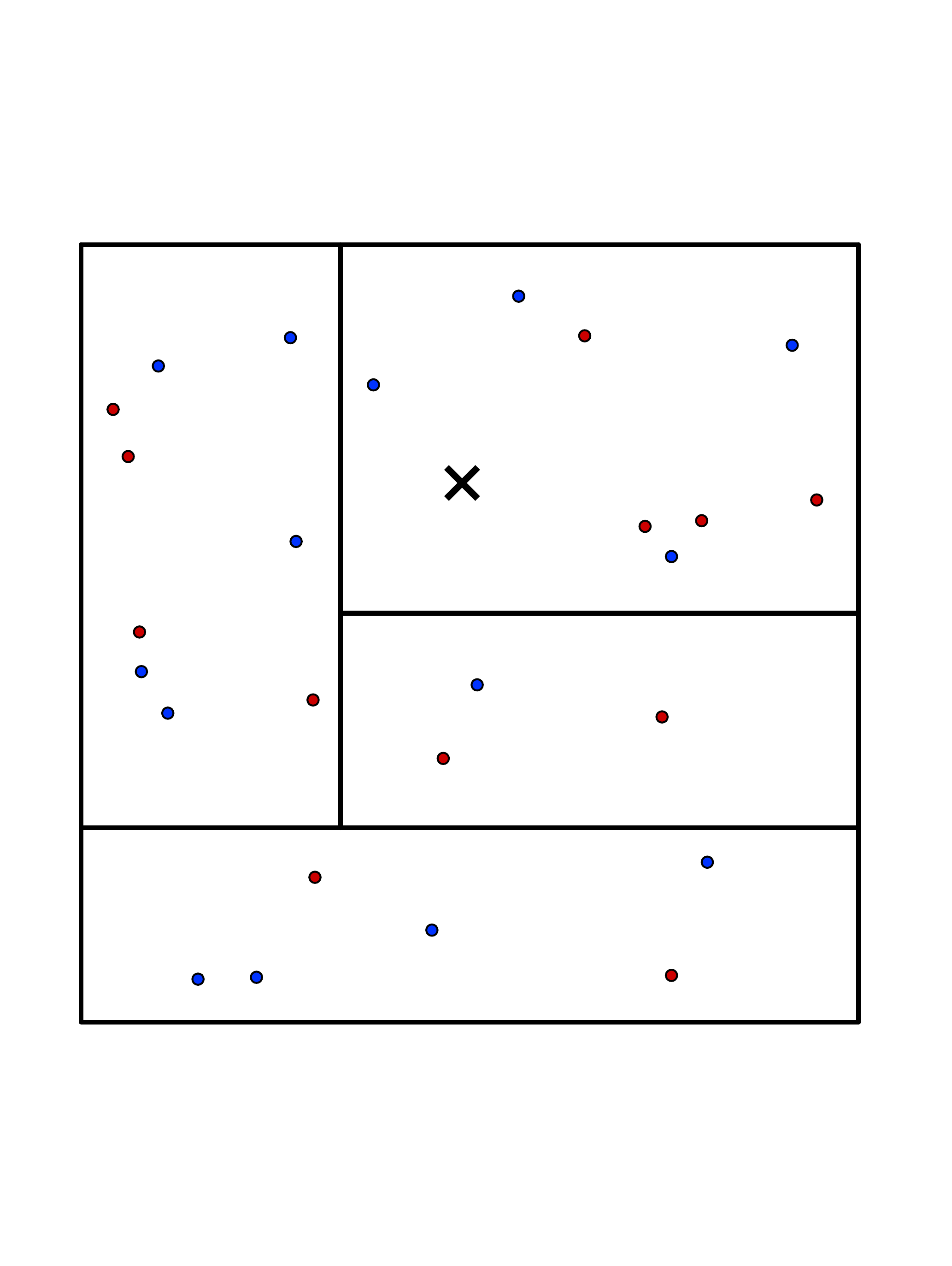} \\
\parbox{0.4\textwidth}{Euclidean neighborhood, \\ for KNN matching.} & &
\parbox{0.4\textwidth}{Tree-based neighborhood.}
\end{tabular}
\end{center}

One way to interpret a tree is that it is an
alternative to kernel regression.  Within each tree, the prediction for a leaf is simply the sample average outcome within the leaf.  Thus, we can think of the leaf as defining the set of nearest neighbors for a given target observation in a leaf, and the estimator from a single regression tree is a matching estimator with non-standard ways of selecting the nearest neighbor to a target point.  In particular, the neighborhoods will prioritize some covariates over others in determining which observations qualify as ``nearby.''   The figure illustrates the difference between  kernel regression and a tree-based matching algorithm for the case of two covariates.  Kernel regression will create a neighborhood around a target observation based on the Euclidean distance to each point, while tree-based neighborhoods will be rectangles.  In addition, a target observation may not be in the center of a rectangle.  Thus, a single tree is generally not the best way to predict outcomes for any given test point $x$.  When a prediction tailored to a specific target observation is desired, generalizations of tree-based methods can be used.

For better estimates of $\mu(x)$, {\it random forests} (\citet{Breiman2001}) build on the regression tree algorithm. A key issue random forests address is that the estimated regression function given a tree is discontinuous with substantial jumps, more than one might like. Random forests induce smoothness by averaging over a large number of trees. These trees differ from each other in two ways. First, each tree is based not on the original sample, but on a bootstrap sample (known as {\it bagging} (\citet{breiman1996bagging})) or alternatively on a subsample of the data.   Second, the splits at each stage are not optimized over all possible covariates, but over a random subset of the covariates, changing every split. These two modifications lead to sufficient variation in the trees that the average is relatively smooth (although still discontinuous), and, more importantly, has better predictive power than a single tree. 

Random forests have become very popular methods. A key attraction is that they require relatively little tuning and have great performance out-of-the-box compared to more complex methods such as deep learning neural networks. 
Random forests and regression trees are particularly effective in settings with a large number of features that are not related to the outcome, that is, settings with sparsity. The splits will generally ignore those covariates, and as a result the performance will remain strong even in settings with a large number of features.  Indeed, when comparing forests to kernel regression, a reliable way to improve the relative performance of random forests perform  is to add irrelevant covariates that have no predictive power.  These will rapidly degrade the performance of kernel regression, but will not affect random forest nearly as severely because it will largely ignore them
\citep{wager2017estimation}.  

Although the statistical analysis of forests had proved elusive since  Breiman's original work, \citet{wager2017estimation} show that a particular variant of random forests can produce estimates $\hat\mu(x)$ with an asymptotically normal distribution centered on the true value $\mu(x)$, and further, they provide an estimate of the variance of the estimator so that centered confidence intervals can be constructed.  The variant they study uses subsampling rather than bagging; and further, each tree is built using two disjoint subsamples, one used to define the tree, and the second used to estimate sample means for each leaf.  This {\it honest} estimation is crucial for the asymptotic analysis.

Random forests can be connected to traditional econometric methods in several ways.  Returning to the kernel regression comparison, since each tree is a form of matching estimator, the forest is an average of matching estimators.  By averaging over trees, the prediction for each point will be centered on the test point (except near boundaries of the covariate space). However, the forest prioritizes more important covariates for selecting matches in a data-driven way.  
Another way to interpret random forests (e.g. \citet{athey2017generalized}), is that they generate weighting functions analogous to kernel weighting functions.  For example, a kernel regression makes a prediction at a point $x$ by averaging nearby points, but weighting closer points more heavily.  A random forest, by averaging over many trees, will include nearby points more often than distant points.  We can formally derive a weighting function for a given test point by counting the share of trees where a particular observation is in the same leaf as a test point. Then, random forest predictions can be written as
\begin{equation}
\label{eq:rf-kernel}
\hat{\mu}_{\text{rf}}(x) = \sum_{i=1}^n \alpha_i(x) \, Y_i, \ \ \sum_{i = 1}^n \alpha_i(x) = 1, \ \ \alpha_i(x) \geq 0,
\end{equation}
where the weights $\alpha_i(x)$ encode the weight given by the forest to the
$i$-th training example when predicting at $x$.  The difference between typical kernel weighting functions and forest-based weighting functions is that the forest weights are adaptive; if a covariate has little effect, it will not be used in splitting leaves, and thus the weighting function will not be very sensitive to distance along that covariate.

\begin{center}
\begin{tabular}{cccc}
\includegraphics[width = 0.25\textwidth, trim=15mm 40mm 20mm 40mm, clip = TRUE]{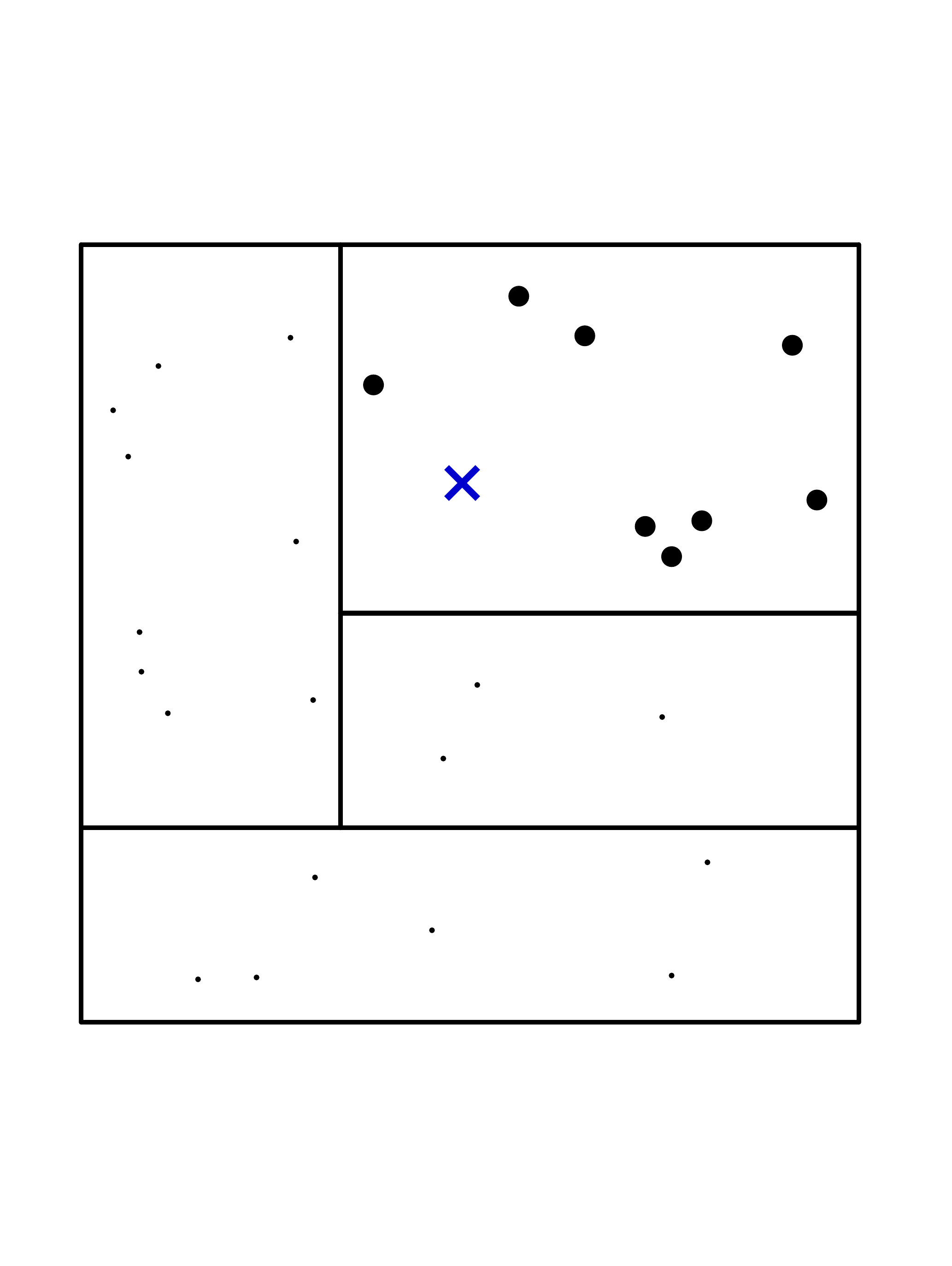} &
\includegraphics[width = 0.25\textwidth, trim=15mm 40mm 20mm 40mm, clip = TRUE]{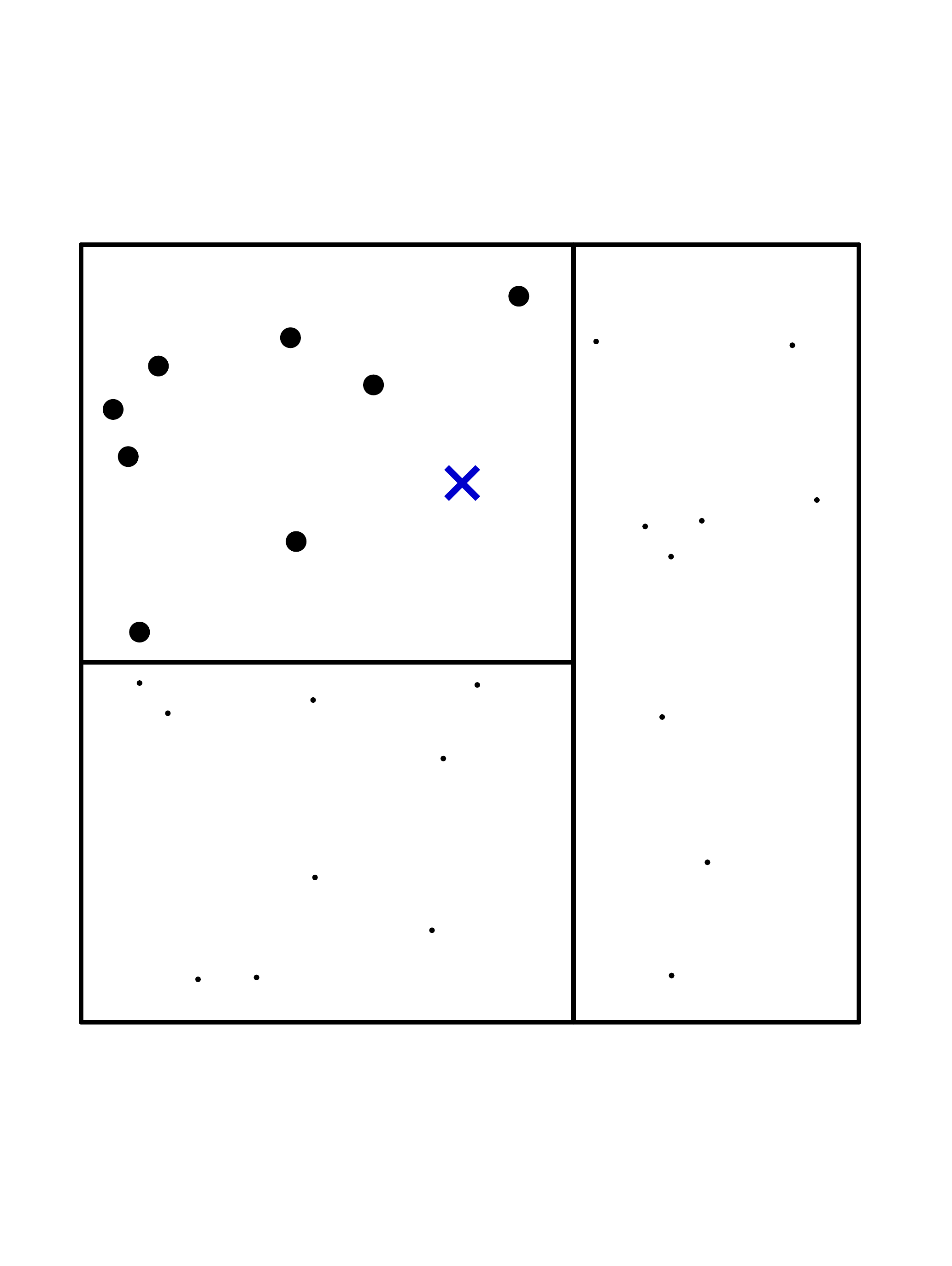} &
\includegraphics[width = 0.25\textwidth, trim=15mm 40mm 20mm 40mm, clip = TRUE]{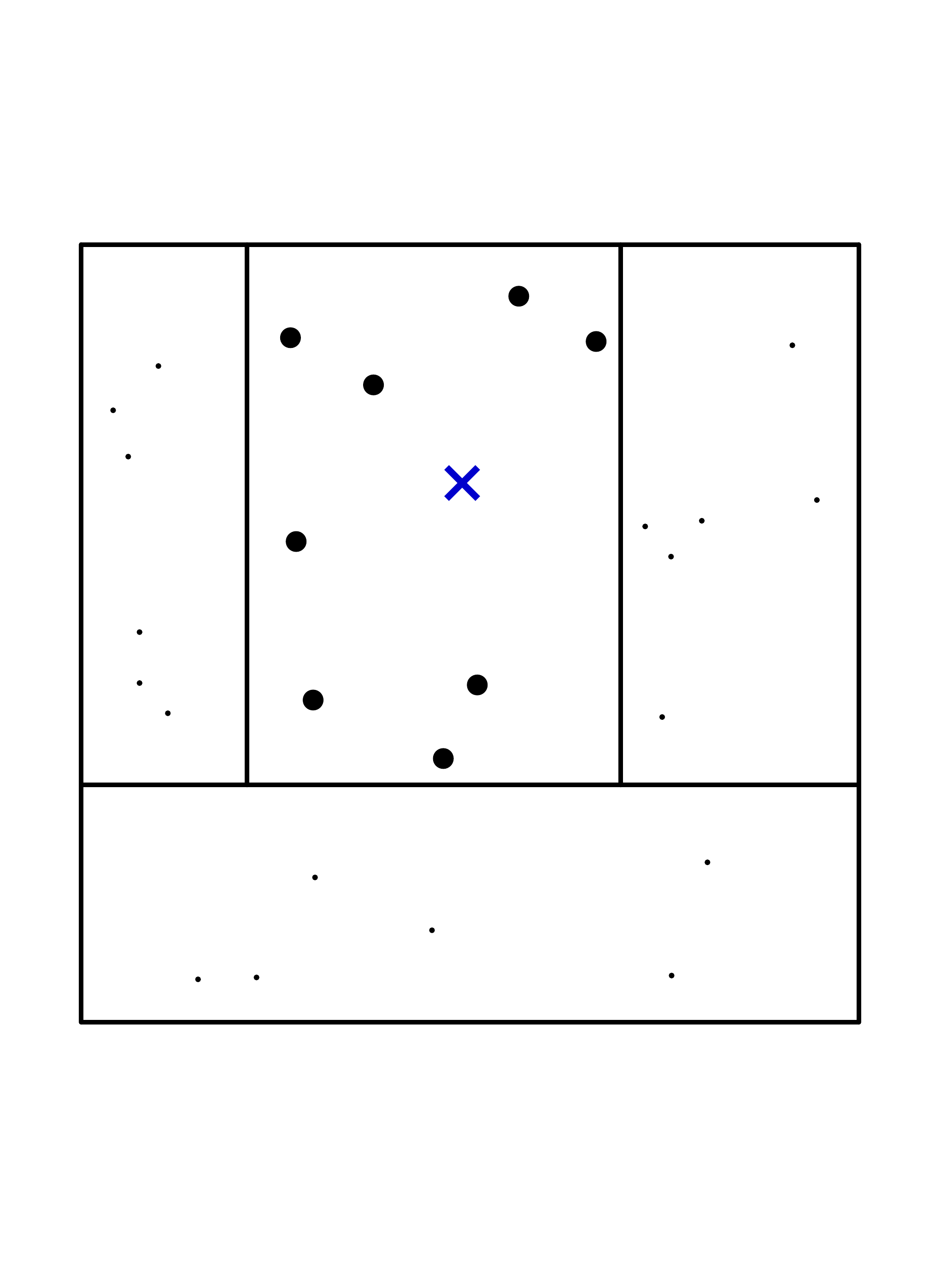} &
\parbox[b][0.25\textwidth][c]{0.1\textwidth}{}
\end{tabular}
Different Trees in Random Forest Generating Weights for Test Point X
\vspace{0.5\baselineskip}

\begin{tabular}{cc}
\parbox[b][0.33\textwidth][c]{0.15\textwidth}{} &
\includegraphics[width = 0.33\textwidth, trim=15mm 40mm 20mm 40mm, clip = TRUE]{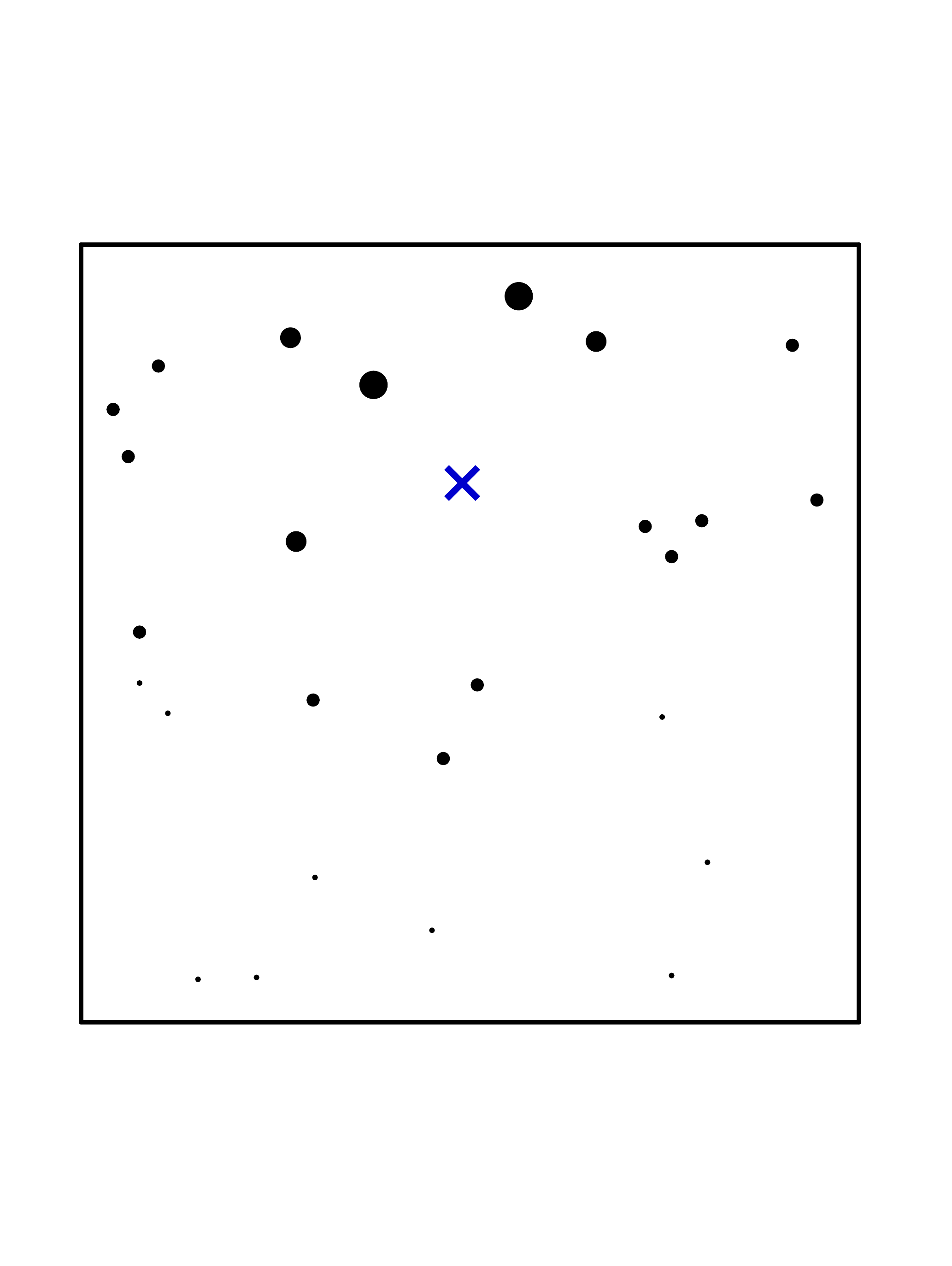}
\end{tabular}

The Kernel Based on Share of Trees in Same Leaf as Test Point X
\end{center}

Recently random forests  have been extended to settings where the interest is in causal effects, either average or unit-level causal effects (\citet{wager2017estimation}), as well as for estimating parameters in general economic models that can be estimated with maximum likelihood or Generalized Method of Moments (GMM, \citet{athey2017generalized}). In the latter case, the interpretation of the forest as creating a weighting function is operationalized; the new {\it generalized random forest} algorithm operates in two steps.  First, a forest is constructed, and second, a GMM model is estimated for each test point, where points that are nearby in the sense of frequently occuring in the same leaf as the test point are weighted more heavily in estimation.  With an appropriate version of honest estimation, these forests produce parameter estimates with an asymptotically normal distribution.  Generalized random forests can be thought of as a generalization of local maximum likelihood, introduced by \citet{tibshirani1987local}, but where kernel weighting functions are used to weight nearby observations more heavily than observations distant from a particular test point.

\begin{figure}[h]
\begin{center}
\begin{tabular}{cc}
\includegraphics[width=0.45\textwidth]{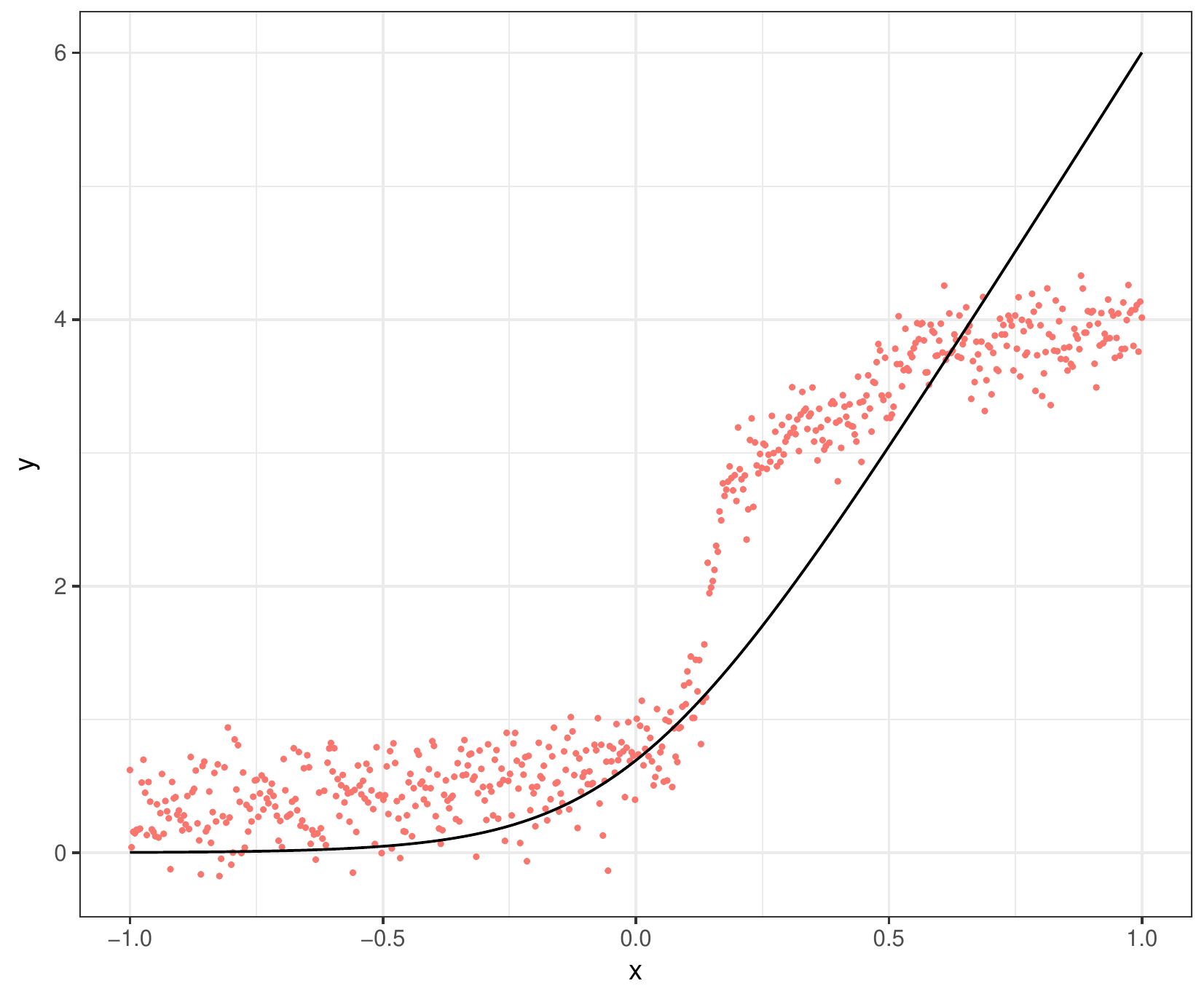} &
\includegraphics[width=0.45\textwidth]{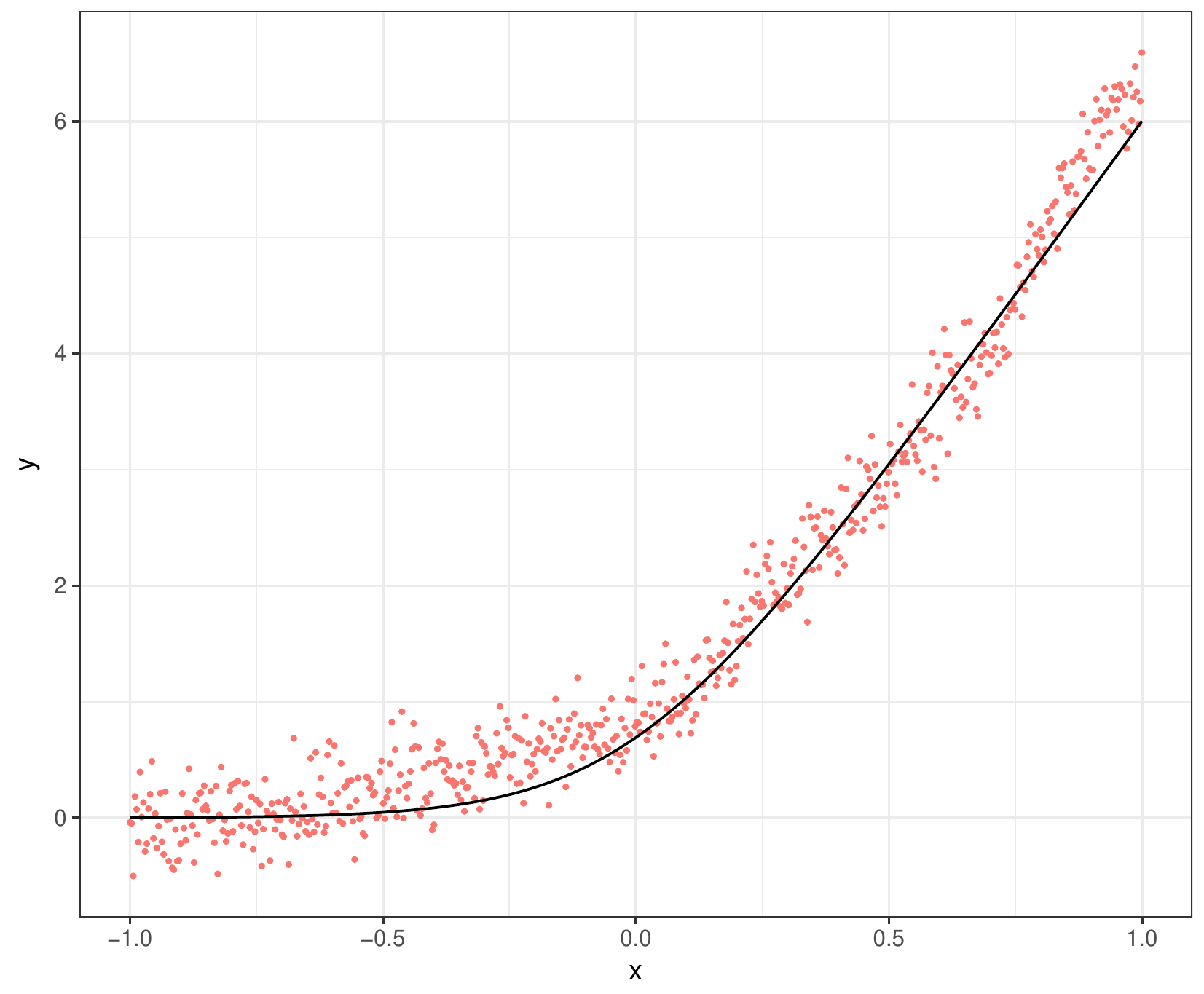} \\
Random forest & Local linear forest
\end{tabular}
\caption{Predictions from random forests and local linear forests on 600 test points. Training and test data were simulated from $Y_i = \text{log}\left(1+e^{6X_{i1}}\right) + \epsilon_i,  \epsilon_i \sim \mathcal{N}(0, \, 20) $ with $X$ having dimension $d=20$ (19 covariates are irrelevant) and errors $\epsilon \sim N(0,20)$. Forests were trained on n = 600 training points using the R package GRF, and tuned via cross-validation. Here the true conditional mean signal $\mu(x)$ is in black, and predictions are shown in red. }
\label{boundary-bias-sims}
\end{center}
\end{figure}

A weakness of forests is that they are not very efficient at capturing linear or quadratic effects, or at exploiting smoothness of the underlying data generating process.  In addition, near the boundaries of the covariate space, they are likely to have bias, because the leaves of the component trees of the random forest cannot be centered on points near the boundary.  Traditional econometrics encounters this boundary bias problem in analyses of regression discontinuity designs where, 
for example, geographical boundaries of school districts or test score cutoffs determine eligibility for schools or programs (\citet{imbens2008regression}).  The solution proposed in the econometrics literature, for example in the matching literature (\citet{abadie2011bias}) is to use local linear regression, which is a regression with nearby points weighted more heavily.  Suppose that the conditional mean function is increasing as it approaches the boundary. Then the local linear regression corrects for the fact that at a test point near the boundary, most sample points lie in a region with lower conditional mean than the conditional mean at the boundary.  \citet{friedberg2018local} extends the generalized random forest framework to local linear forests, which are constructed by running a regression weighted by the weighting function derived from a forest.  In their simplest form, local linear forests just take the forest weights $\alpha_i(x)$, and use them for
local regression:
\begin{equation}\label{loss}
 (\hat{\mu}(x),\hat{\theta}(x)) = \text{argmin}_{\mu,\theta} \left\{\sum_{i=1}^n \alpha_i(x) (Y_i - \mu(x) - (X_i-x)\theta(x) )^2 + \lambda ||\theta(x)||_2^2\right\}.
\end{equation}
Performance can be improved by modifying the tree construction to incorporate a regression correction; in essence, splits are optimized for predicting residuals from a local regression.  This algorithm performs better than traditional forests in settings where a regression can capture broad patterns in the conditional mean function such as monotonicity or a quadratic structure, and again, asymptotic normality is established.  Figure \ref{boundary-bias-sims}, from \citet{friedberg2018local}, illustrates how local linear forests can improve on regular random forests; by fitting local linear regressions with a random-forest estimated kernel, the resulting predictions can match a simple polynomial function even in relatively small data sets.  In contrast, a forest tends to have bias, particularly near boundaries, and in small data sets will have more of a step function shape. Although the figure shows the impact in a single dimension, an advantage of the forest over a kernel is that these corrections can occur in multiple dimensions, while still allowing the traditional advantages of a forest in uncovering more complex interactions among covariates.

\subsection{Deep Learning and Neural Nets}

Neural networks and related deep learning methods are another general and flexible approach to estimating regression functions. They have been found to be very succesful in complex settings, with extremely large number of features. However, in practice these methods require a substantial amount of tuning in order to work well for a given application, relative to methods such as random forests. Neural networks were studied in the econometric literature in the 1990s, but did not catch on at the time
 (see \citet{white1992artificial,  hornik1989multilayer, white1992artificial}).

Let us consider a simple example. Given $K$ covariates/features $X_{ik}$, we model $K_1$ latent/unobserved variables $Z_{ik}$ ({\it hidden nodes}) that are linear in the original covariates:
\[ Z^{(1)}_{ik}=\sum_{j=1}^K \beta^{(1)}_{kj} X_{ij}, \ \ {\rm for}\ k=1,\ldots,K_1.\]
We then modify these linear combinations using a simple nonlinear transformation,
e.g.,  a sigmoid function
\[ g(z)=(1+\exp(-z))^{-1},\]
or a rectified linear function
\[ g(z)=z{\bf 1}_{z>0},\]
and then
model the outcome as a linear function of  this nonlinear transformation of these hidden nodes plus noise:
\[ Y_i=\sum_{k=1}^{K_1} \beta^{(2)}_k g\left(Z^{(1)}_{ik}\right)+\varepsilon_i.\] This is a neural network with a single hidden layer with $K_1$ hidden nodes. The transformation $g(\cdot)$ introduces nonlinearities in the model. Even with this single layer, with many nodes one can approximate arbitrarily well a  rich set of smooth functions.

 It may be tempting to fit this into a standard framework and
interpret this model simply as a complex, but fully parametric, specification for the potentially nonlinear conditional expectation of $Y_i$ given $X_i$:
\[ \mme[Y_i|X_i=x]=\sum_{k'=1}^{K_1}\beta^{(2)}_{k'}g\left(\sum_{k=1}^K\beta^{(1)}_{k'k}X_{ik}\right).\]
Given this interpretation, we can estimate the unknown parameters using nonlinear least squares. We could then derive the properties of the least squares estimators, and functions thereof, under standard regularity conditions. However, this interpretation of a neural net as a standard nonlinear model would be missing the point, for four reasons. First, it is likely that the asymptotic distributions for the parameter estimates would be poor approximations to the actual sampling distributions. Second, the  estimators for the parameters would be poorly behaved, with likely substantial collinearity without careful regularization. Third, and more important, these properties are not of intrinsic interest. We are interested in the properties of the predictions from these specifications, and these can be quite attractive even if the properties of the paramater estimates are not. Fourth, we can make these models much more flexible, and at the same time, make the properties of the corresponding least squares estimators of the parameters substantially less tractable and attractive, by adding layers to the neural network. A second layer of hidden nodes would have representations that are linear in the same transformation $g(\cdot)$ of linear combinations of the first layer of hidden nodes:
\[ Z^{(2)}_{ik}=\sum_{j=1}^{K_1} \beta^{(2)}_{kj} g\left(Z^{(1)}_{ij}\right), \ \ {\rm for}\ k=1,\ldots,K_2,\]
with the outcome now a function of the second layer of hidden nodes,
\[ Y_i=\sum_{k=1}^{K_2} \beta^{(3)}_k g\left(Z^{(2)}_{ik}\right)+\varepsilon_i.\]
 The depth of the network substantially increases the flexibility in practice, even if with a single layer and many nodes we can already approximate a very rich set of functions.
 Asymptotic properties for multilayer networks have recently been established in \citet{farrell2018deep}.
  In applications researchers have used models with many layers, {\it e.g.}, ten or more, and millions of parameters:
 \begin{quote}
``We observe  that shallow models [models with few layers] in this context overfit at  around 20 millions parameters while deep ones can benefit from having over 60 million. This suggests that using a deep model  expresses a useful preference over the space of functions the model can learn.''
\citet{lecun2015deep})\end{quote}

   In  cases with multiple hidden layers and many hidden nodes one needs to carefully regularize the parameter estimation, possibly through a penalty term that is proportional to the sum of the squared coefficients in the linear parts of the model.
The architecture of the networks is also important. It is possible, as in the specification above, to have the hidden nodes at a particular layer be a linear function of all the hidden nodes of the previous layer, or restrict them to a subset based on substantive considerations ({\it e.g.,} proximity of covariates in some metric, such as location of pixels in a picture). Such {\it convolutional} networks have been very succesful, but require even more careful tuning (\citet{krizhevsky2012imagenet}).

Estimation of the parameters of the network is based on approximately minimizing the
sum of the squared residuals,
 plus a penalty term that depends on the complexity of the model. This minimization problem is challenging, especially in settings with multiple hidden layers. The algorithms of choice use the {\it back-propagation}  algorithm and variations thereon (\citet{rumelhart1986learning})  to calculate the exact  derivatives with respect to the parameters of the unit-level terms in the objective function. This algorithm  exploits in a clever way the hierarchical structure of the layers, and the fact that each parameter enters only in a single layer. The algorithms then use stochastic gradient descent (\citet{friedman2002stochastic, bottou1998online, bottou2012stochastic}) described in Section \ref{section:scale} as a computationally efficient method for finding the approximate optimum.

\subsection{Boosting}

{\it Boosting} is a general purpose technique to improve the performance of simple supervised learning methods. See
\citet{schapire2012boosting} for a detailed discussion. Let us say we are interested in prediction of an outcome given a substantial number of features. Suppose we have a very simple algorithm for prediction, a simple {\it  base learner}. For example, we could have a regression tree with  three leaves, that is, a regression tree based on a two splits, where we estimate the regression function as the average outcome in the corresponding leaf. Such an algorithm on its own would not lead to a very attractive predictor in terms of predictive performance because it uses at most  two of the many possible features. Boosting improves this {\it base learner} in the following way. Take for all units in the training sample the residual from the prediction based on the simple three-leaf tree model, $Y_i-\hat Y_i^{(1)}$. Now we apply the same {base learner} (here the two split regression tree) with the residuals as the outcome of interest (and with the same set of original features).
 Let $\hat Y^{(2)}_i$ denote the prediction from combining the first and second steps. 
  Given this new tree we can calculate the new residual, $Y_i-\hat Y_i^{(2)}$. We can then repeat these step, using the new residual as the outcome and again constructing a two split regression tree. We can do this many times, and get a prediction based on re-estimating the basic model many times on the updated residuals.

If we base our boosting algorithm on  a regression tree with $L$ splits, it turns out that the resulting predictor can approximate any regression function that can be written as the sum of functions of $L$ of the original features at a time. So, with $L=1$, we can approximate any function that is additive in the features, and with $L=2$ we can approximate any function that is additive in functions of the original features that allow for general second order effects.

Boosting can also be applied using  base learners other than regression trees. The key is to choose a base learner that is easy to apply many times without running into computational problems.


\section{Supervised Learning for Classification Problems}
\label{section:supervised_classification}

Classification problems are the focus of the other main branch of the supervised learning literature. The problem is, given a set of observations on a vector of features $X_i$, and a label $Y_i$ (an unordered discrete outcome), the goal is a function that assigns new units, on the basis of their features, to one of the labels. This is very closely related to discrete choice analysis in econometrics, where researchers specify statistical models that imply a probability that the outcome takes on a particular value, conditional on the covariates/features.
Given such an probability it is of course straightforward to predict a unique label, namely the one with the highest probability. However, there are differences between the two approaches. An important one is that in the classification literature the focus is often  solely on the classification, the choice of a single label. One can classify given a probability for each label, but one does not need such a probability to do the classification. Many of the classification methods do not, in fact, first estimate a probability for each label, and so are not directly relevant in settings where such a probability is required. A practical difference is that the classification literature has often focused on settings where ultimately the covariates allow one to assign the label with almost complete certainty, as opposed to settings where even the best methods have high error rates. 

The classic example is that of digit recognition. Based on a picture, coded as a set of say 16 or 256 black and white pixels, the challenge is to classify the image as corresponding to one of the ten digits from 0 to 9.  Here ML methods have been spectacularly successful. Support Vector Machines
(SVMs \citet{cortes1995support}) greatly outperformed other methods in the nineties. More recently 
deep convolutional neural networks (\citet{krizhevsky2012imagenet}) have improved error rates even further.

\subsection{Classification Trees and Forests}

Trees and random forests are easily modified from a focus on estimation of regression functions to classification tasks. See \citet{breiman1984classification} for a general discussion. Again we start by splitting the sample into two leaves, based on a single covariate exceeding or not a threshold. We optimize the split over the choice of covariate and the threshold. The difference between the regression case and the classification case is in the objective function that measures the improvement from a particular split.
In classification problems this is called the {\it impurity} function. It measures, as a function of the shares of units in a given leaf with a particular label, how {\it impure} that particular leaf is. If there are only two labels, we could simply assign the labels the numbers zero and one, interpret the problem as one of estimating the conditional mean and use the average squared residual as the impurity function. That does not generalize naturally to the multi-label case. Instead a more common impurity function, as a function of the $M$ shares $p_1,\ldots,p_M$ is the Gini impurity, 
\[ I(p_1,\ldots,p_M)=-\sum_{m=1}^M p_m\ln(p_m).\]
This impurity function is minimized if the leaf is {\it pure}, meaning that all units in that leaf have the same label, and is maximized if the shares are all equal  to $1/M$.
The regularization typically works again through a penalty term on the number of leaves in the tree.
The same extension from a single tree to a random forest that was discussed for the regression case works for the classification case.

\subsection{Support Vector Machines and Kernels}

{\it Support Vector Machines} (SVMs, \citet{vapnik1998statistical, scholkopf2001learning}) are another flexible set of methods for classification analyses. SVMs can also extended to regression settings, but are more naturally introduced in a classification context, and for simplicity we focus on the case with two possible labels. Suppose we have a set with $N$ observations on a $K$-dimensional vector of features $X_i$ and a binary label $Y_i\in\{-1,1\}$ (we could use 0/1 labels but using -1/1 is more convenient). Given a $K$-vector  of {\it weights} $\omega$ (what we would typically call the parameters) and a constant $b$ (often called the {\it bias} in the SVM literature), define the hyperplane $x\in\mathbb{R}$ such that $\omega^\top x+b=0$. We can think of this hyperplane defining a binary classifier ${\rm sgn}(\omega^\top X_i+b)$, with units $i$ with $\omega^\top x+b\geq 0$ classified as  1 and units with  $\omega^\top x+b<0$ classified as -1.
Now consider for each hyperplane (that is, for each pair $(\omega,b)$) the number of classification errors in the sample. If we are very fortunate there would be some hyperplanes with no classification errors. In that case there are typically many such hyperplanes, and we choose the one that maximizes the distance to the closest units. There will typically be a small set of units that have the same distance to the hyperplane (the same {\it margin}). These are called the {\it support vectors}.

We can write this as an optimization problem as
\[(\hat\omega,\hat b)=\arg\min_{\omega,b} \left\|\omega\right\|^2,\hskip1cm   {\rm  subject\ to\ } Y_i ( \omega^\top X_i +b ) \geq 1 ,\ \ {\rm for\ all\ } i=1,\ldots,N.\]
with classifier
\[ {\rm sgn}(\hat\omega^\top X_i+\hat b).\]
Note that if there is a hyperplane with no classification errors, a standard logit model would not have a maximum likelihood estimator: the argmax of the likelihood function would diverge.

We can also write this problem in terms of the Lagrangian, with $\alpha_i$ the Lagrangian multiplier for the restriction $Y_i ( \omega^\top X_i +b ) \geq 1 $,
\[\min_{\alpha, \omega,b}\left\{\frac{1}{2} \|\omega\|^2-\sum_{i=1}^N \alpha_i(Y_i(\omega^\top X_i+b)-1)\right\},\hskip1cm {\rm subject\ to}\ 0\leq \alpha_i.\]
After concentrating out the weights $\omega$ this is  equivalent to 
\[ \max_{\alpha} \left\{\sum_{i=1}^N \alpha_i-\frac{1}{2}\sum_{i=1}^N \sum_{j=1}^N \alpha_i\alpha_j Y_i Y_i X_i^\top X_j\right\},\hskip1cm {\rm subject\ to\ }0\leq \alpha_i,\ \ \sum_{i=1}^N \alpha_i Y_i=0,\]
where $\hat b$ solves $\sum_i\hat \alpha_i(Y_i(X_i^\top\omega+\hat b)-1)=0$,
with classifier
\[ f(x)={\rm sgn}\left(\hat b+\sum_{i=1}^N Y_i\hat \alpha_i X_i^\top x\right),\]

 In practice, of course, we are typically in a situation where there exists no hyperplane without  classification errors. In that case there is no solution as the $\alpha_i$ diverge for some $i$. We can modify the classifier by adding the constraint that the $\alpha_i\leq C$. \citet{scholkopf2001learning} recommend setting $C=10 N$.

   This is still a linear problem, differing from a logistic regression only in terms of the loss function. Units far away from the hyperplane do not affect the estimator as much in the SVM approach as they do in a logistic regression, leading to more robust estimates.
   However, the real power from the SVM approach is in the nonlinear case. We can think of that in terms of constructing a number of functions of the original covariates, $\phi(X_i)$, and then finding the optimal hyperplane in the transformed feature space. However, because the features enter only through the inner product $X_i^\top X_j$, it is possible to skip the step of specifying the transformations $\phi(\cdot)$, and instead directly write the classifier in terms of a kernel $K(x,z)$,
   through
 \[ \max_{\alpha} \sum_{i=1}^N \left\{\alpha_i-\frac{1}{2}\sum_{i=1}^N \sum_{j=1}^N \alpha_i\alpha_j Y_i Y_i K(X_i,X_j)\right\},\hskip0.5cm {\rm subject\ to\ }0\leq \alpha_i\leq C,\ \ \sum_{i=1}^N \alpha_i Y_i=0,\] 
where $\hat b$ solves $\sum_i\hat \alpha_i(Y_i(X_i^\top\omega+\hat b)-1)=0$,
with classifier
\[ f(x)={\rm sgn}\left(\sum_{i=1}^N Y_i\alpha_i K(X_i,x)+b\right).\]
Common choices for the kernel are $k_h(x,z)=\exp(-(x-z)^\top(x-z)/h)$, or $k_{\kappa,\Theta}(x,z)={\rm tanh}
(\kappa (x-z)^\top(x-z)+\Theta)$. The parameters of the kernel, capturing the amount of smoothing, are typically chosen through crossvalidation.

\section{Unsupervised Learning}
\label{section:unsupervised}

A second major topic in the ML literature is unsupervised learning. In this case we have a number of cases without labels. We can think of that as having a number of observations on covariates without an outcome. We may be interested in partitioning the sample into a number of subsamples or {\it clusters}, or in estimating the joint distribution of these variables.

\subsection{K-means Clustering}

 Here the goal is given a set of observations on features $X_i$, to partition the feature space into a number of subspaces. These clusters may be used to to create new features, based on subspace membership. 
For example, we may wish to use the partioning to estimate parsimonious models within each of the subspaces. We may also wish to use cluster membership as a way to organize the sample into types of units that may receive different exposures to treatments.
This is an unusual problem, in the sense that there is no natural benchmark to assess whether a particular solution is a good one relative to some other one. 
A closely related approach that is more traditional in the econometrics and statistics literatures is {\it mixture} models, where the distribution that generated the sample is modelled as a mixture of different distributions. The mixture components are similar in nature to the clusters.

A key method is the k-means algorithm (\citet{hartigan1979algorithm, alpaydin2009introduction}).
 Consider the case where we wish to partition the feature space into $K$ subspaces or clusters. We wish to choose centroids $b_1,\ldots, b_K$, and then assign units to the cluster based on their proximity to the centroids. 
The basic algorithm works as follows. We start with a set of $K$ centroids, $b_1,\ldots,b_K$, elements of the feature space, and sufficiently spread out over this space. Given a set of centroids, assign each unit to the cluster that minimizes the distance between the unit and the centroid of the cluster:
\[ C_i=\arg\min_{c\in\{1,\ldots,K\}} \left\| X_i-b_c\right\|^2.\]
Then update the centroids as the average of the $X_i$ in each of the clusters:
\[ b_c=\sum_{i:C_i=c} X_i\biggl/\sum_{i:C_i=c} 1.\]
Repeatedly interate between the two steps. 
Choosing the number of clusters $K$ is difficult because there is no direct cross-validation method to assess the performance of one value versus the other. Often this number is chosen on substantive grounds rather than in a data-driven way.

There are a large number of alternative unsupervised methods, including topic models, which we discuss further below in the section about text.  Unsupervised variants of neural nets are particularly popular for images and videos.

\subsection{Generative Adverserial Networks}

Now let us consider the problem of estimation of a joint distribution, given observations on $X_i$ for a random sample of units. A recent ML approach to this is Generative Adverserial Networks (GANs,
\citet{arjovsky2017towards, goodfellow2014generative}).
The idea is to find an algorithm to generate data that look like the sample $X_1,\ldots,X_N$. 
A key insight is that  there is an effective way of assessing whether the algorithm is succesful that is like a Turing test. If we have a succesful algorithm we should not be able to tell whether data were generated by the algorithm, or came from the original sample. Hence we can assess the algorithm by training a classifier on data from the algorithm and a subsample from the original data. If the algorithm is succesful the classifier cannot be succesfully classifying the data as coming from the original data or the algorithm. The GAN then uses the relative success of the classification algorithm to improve the algorithm that generates the data, in effect pitting the classification algorithm against the generating algorithm.

This type of algorithm may also be an effective way of choosing simulation designs intended to mimic real world data.

\section{Machine Learning and Causal Inference}

An important difference between much of the econometrics literature and the machine learning literature is that the econometrics literature is often focused on questions beyond simple prediction. In many, arguably most, cases, researchers are interested in average treatment effects or other causal or structural parameters (see \citet{abadie2018econometric} and \citet{imbenswooldridge} for surveys).
Covariates that are of limited importance  for prediction may still play an important role in estimating such structural parameters. 

\subsection{Average Treatment Effects}
A canonical problem is that of estimating average treatment effects under unconfoundedness (\citet{rosenbaum1983central, imbens2015causal}). Given data on an outcome $Y_i$, a binary treatment $W_i$, and a vector of covariates or features $X_i$, a common estimand, the Average Treatment Effect (ATE) is defined as $\tau=\mathbb{E}[Y_i(1)-Y_i(0)]$, where $Y_i(w)$ is the potential outcome unit $i$ would have experienced if their treatment assignment had been $w$.  Under the unconfoundedness assumption, 
 which ensures that potential outcomes are independent of the treatment assignment conditional on covariates 
\[W_i\ \indep\ \Bigl(Y_i(0),Y_i(1)\Bigr)\  \Bigl|\ X_i,\]
 the ATE is identified.  The ATE can be characterized  in an number of different ways as a functional of the joint distribution of $(W_i,X_i,Y_i)$. Three important ones are  $(i)$ as the covariate-adjusted difference between the two treatment groups, $(ii)$  as a weighted average of the outcomes, and $(iii)$ in terms of the influence or efficient score function.
\begin{equation} 
\begin{aligned}
 \tau  &=\mme\left[
\mu(1,X_i)-\mu(0,X_i)
\right] \\
  &=\mme\left[\frac{Y_i W_i}{e(X_i)}-
\frac{Y_i (1-W_i)}{1-e(X_i)}\right] \\
&=\mme\left[\frac{(Y_i-\mu(1,X_i)) W_i}{e(X_i)}-
\frac{(Y_i-\mu(0,X_i)) (1-W_i)}{1-e(X_i)}\right] +\mme\left[\mu(1,X_i)-\mu(0,X_i)\right],
\end{aligned}
\label{eq:ATE}
\end{equation}
where
\[ \mu(w,x)=\mme[Y_i|W_i=w,X_i=x],\hskip1cm {\rm and}\ \ \ e(x)=\mme[W_i|X_i=x].\]
One can estimate the average treatment effect  by using the first representation by estimating the conditional outcome expectations
$\mu(\cdot)$, using the second representation  by estimating the propensity score $e(\cdot)$, or using the third representation and estimating both the conditional outcome expectation and the propensity score. Given a particular choice of representation, there is the question of the appropriate estimator for the particular conditional expectations that enter into that representation. For example, if we wish to use the first representation, and want to consider linear models, it may seem natural to use LASSO or subset selection. However, as illustrated in \citet{belloni2014jep}, such a strategy could have very poor properties. The set of features that is optimal for inclusion when the objective is  estimating $\mu(\cdot)$ is not necessarily optimal for estimating $\tau$. The reason is that omitting from the regression covariates that are highly correlated with the treatment $W_i$ can introduce substantial biases even if their correlation with the outcome is only modest.  Thus, optimizing model selection solely for predicting outcomes is not the best approach.
 \citet{belloni2014jep} propose using a covariate selection method that selects both covariates that are predictive of the outcome and covariates that are predictive of the treatment, and show that this substantially improves the properties of the corresponding estimator for $\tau$.
 
More recent methods focus on combinations of estimating both the conditional outcome expectations $\mu(\cdot)$  and the propensity score 
$e(\cdot)$ flexibly and combining them in doubly robust methods (\citet{robins1, chernozhukov2016double, chernozhukov2016locally}), and methods that combine estimating the conditional outcome expectations $\mu(\cdot)$ with covariate balancing (\citet{atheyimbenswager}).  Covariate balancing is inspired by another common approach in ML, which  frame data analysis as an optimization problem.  Here, instead of trying to estimate a primitive object, the propensity score $e(\cdot)$, the optimization procedure directly optimizes weights for the observations that lead to the same mean values of covariates in the treatment and control group (\citet{zubizarreta2015stable}).  This approach allows for efficient estimation of average treatment effects even when the propensity score is too complex to estimate well.  Because traditional propensity score weighting entails dividing by the estimated propensity score, instability in propensity score estimation can lead to high variability in estimates for the average treatment effect.  Further, in an environment with many potential confounders, estimating the propensity score using regularization may lead to the omission of weak confounders that still contribute to bias.  Directly optimizing for balancing weights can be more effective in environments with many weak confounders.  

The case of estimating average treatment effects under unconfoundedness is an example of a more general theme from econometrics; typically,
economists prioritize precise estimates of causal effects above predictive power (see \citet{athey2017beyond, athey2018impact} for further elaboration of this point).  In instrumental variables models, it is common that goodness of fit falls by a substantial amount between an ordinary least squares regression and the second stage of a two-stage least squares model. However, the instrumental variables estimate of causal effects can be used to answer questions of economic interest, and so the loss of predictive power is viewed as less important. 

\subsection{Orthogonalization and Cross-Fitting}

A theme that has emerged across multiple distinct applications of machine learning to parameter estimation is that both practical performance and theoretical guarantees can be improved by using two simple techniques, both involving nuisance parameters that are estimated using machine learning.  These can be illustrated through the lens of estimation of average treatment effects.  Building from the third representation in (\ref{eq:ATE}), we can define the influence function of each observation as follows:
\[\psi(y,w,x) = \mu(1,x)-\mu(0,x) + \frac{w}{e(x)} (y - \mu(1,x) + \frac{1-w}{1-e(x)} (y - \mu(0,x)),\] 
with $ \Psi_i=\psi(Y_i,W_i,X_i) .$
An estimate of the ATE can be constructed by first constructing estimates $\hat\mu(w,x)$ and $\hat{e}(x)$, and plugging those in to get an estimate  $\hat\Psi_i=\hat\psi(Y_i,W_i,X_i)$ for each observation.  Then, the sample average of $\hat\Psi_i$ is an estimator for the ATE.  This approach is analyzed in \citet{bickel} and  \citet{van2000asymptotic} for the general semiparametric case, and  in \citet{chernozhukov2017double} for the average treatment effect case. 
A key result is that an estimator based on this approach is efficient if the estimators are sufficiently accurate in the following sense: 
\[\mme[(\hat\mu(w,X_i)-\mu(w,X_i))^2]^\frac{1}{2}\mme[(\hat{e}(X_i)-e(X_i))^2]^\frac{1}{2} = o_P \left( N^{-1/2} \right). \]
For example, each nuisance component, $\hat\mu(\cdot)$ and $\hat{e}(\cdot)$, could converge at rate close to $N^{-1/4}$, an order of magnitude slower than the ATE estimate.  This works because $\Psi_i$ makes use of orthogonalization; by construction, errors in estimating the nuisance components are orthogonal to errors in $\Psi_i$.  This idea is more general, and has been exploited in a series of papers, with theoretical analysis in \citet{chernozhukov2018double, chernozhukov2018riesz}, and other applications including \citet{athey2017generalized} for estimating heterogeneous effects in models with unconfoundedness or those that make use of instrumental variables.

A second idea, also exploited in the same series of papers, is that performance can be improved using techniques such as sample splitting, cross-fitting, out-of-bag prediction, and leave-one-out estimation.  All of these techniques have the same final goal: nuisance parameters estimated to construct the influence function $\hat\Psi_i$ for observation $i$ (for the ATE case, $\hat\mu(w,X_i)$ and $\hat{e}(X_i)$) should be estimated without using outcome data about observation $i$.  When random forests are used to estimate the nuisance parameters, this is straightforward, since ``out-of-bag'' predictions (standard in random forest statistical packages) provide the predictions obtained using trees that were constructed without using observation $i$.  When other types of ML models are used to estimate the nuisance parameters, ``cross-fitting'' or sample splitting advocates splitting the data into folds and estimating the nuisance parameters separately on all data except a left-out fold, and then predicting the nuisance paramters in the left-out fold.  When there are as many folds as observations, this is known as leave-one-out estimation.

Although these two issues are helpful in traditional small--data applications, when ML is used to estimate nuisance parameters (because there are many covariates), these issues become much more salient.  Overfitting is more of a concern, and in particular, a single observation $i$ can have a strong effect on the predictions made for covariates $X_i$ when the model is very flexible.  Cross-fitting can solve this problem.  Second, we should expect that with many covariates relative to the number of observations, accurate estimation of nuisance parameters is harder to achieve.  Thus, orthogonalization makes estimation more robust to these errors.

\subsection{Heterogenous Treatment Effects} 

Another place where machine learning can be very useful is in uncovering treatment effect heterogeneity, where we focus on heterogeneity
with respect to observable covariates.  Examples of questions include, which individuals benefit most from a treatment?  For which individuals is the treatment effect positive?  How do treatment effects change with covariates?  Understanding treatment effect heterogeneity can be useful for basic scientific understanding, or for estimating optimal policy assignments; see \citet{athey2017state} for further discussion.

Continuing with the potential outcome notation from the last subsection, we define the conditional average treatment effect (CATE) as $\tau(x)=E[\tau_i|X_i=x]$, where $\tau_i=Y_i(1)-Y_i(0)$ is the treatment effect for individual $i$.  The CATE is identified under the unconfoundedness assumption introduced in the last subsection. Note that $\tau_i$ can not be observed for any unit; this ``fundamental problem of causal inference'' (\citet{holland1986statistics}) is the source
of an apparent difference between estimating heterogeneous treatment effects and predicting outcomes, which are typically observed for each unit.  

We focus here on three types of questions: (i) learning a
low-dimensional representation of treatment effect heterogeneity, and conducting hypothesis tests about this heterogeneity; (ii) learning a flexible (nonparametric) estimate of $\tau(x)$; and (iii) estimating an optimal policy allocating units to either treatment or control on the basis of covariates $x$.  

An important issue in adapting machine learning methods to focus on causal parameters relates to the criterion function used in model selection.  Predictive models typically use a mean squared error (MSE) criterion, $\sum_i (Y_i - \hat\mu(X_i))^2/N$ to evaluate performance. Although the MSE in a held-out test set is a noisy estimate to the population expectation of the MSE in an independent set, the sample average MSE is a good, that is, unbiased, approximation that does not rely on further assumptions (beyond independence of observations), and the standard error of the squared errors in the test set accurately captures the uncertainty in the estimate.  In contrast, consider the problem of estimating the CATE in observational studies.  It would be natural to use as a criterion function the mean squared error of treatment effects, 
$\sum_i(\tau_i - \hat\tau(X_i))^2/N$, where $\hat\tau(x)$ is the estimate of the CATE.  However, this criterion is infeasible, since we do not observe unit-level causal effects.
Further, there is no simple, model-free unbiased estimate of this criterion in observational studies.  For this reason, comparing estimators, and as a result developing regularization strategies, is a
substantially harder challenge in settings where we are interested in structural or causal parameters than in settings where we are interested in predictive performance.

These difficulties in finding effective cross-validation strategies are not always unsurmountable, but they lead to a need to carefully adapt and modify basic regularization methods to address the questions of interest. \citet{athey2016recursive} 
proposes several different possible criterion to use for optimizing splits as well as for cross-validation.  A first insight is that when conducting model selection, it is only necessary to compare models.  The $\tau_i^2$ term (which would be difficult to estimate) cancels out when comparing two estimators, say $\hat\tau'(x)$ and $\hat\tau''(x)$.  The remaining terms are linear in $\tau_i$, and the expected value of $\tau_i$ can be estimated.  If we define what \citet{athey2016recursive}  call the {\it transformed outcome}
$$Y_i^*=W_i \frac{Y_i}{e(X_i)} - (1-W_i) \frac{Y_i}{1-e(X_i)},$$
then $\mme[Y_i^*|X_i]=\mme[\tau_i|X_i]$.  When the propensity score is unknown, it must be estimated, which implies that a criterion based on an estimate of the mean squared error of the CATE will depend on modeling choices. 

\citet{athey2016recursive} build on this insight and propose several different estimators for the relative MSE of estimators for the CATE. They develop a method, which they call {\it causal tree,} for learning a low-dimensional representation of treatment effect heterogeneity, which provides reliable confidence intervals for the parameters it estimates.  The paper builds on regression tree methods, creating a partition of the covariate space and then estimating treatment effects in each element of the partition.  Unlike regression trees optimized for prediction, the splitting rule optimizes for finding splits associated with treatment effect heterogeneity.  In addition, the  method relies on sample splitting; half the data is used to estimate the tree structure, and the other half (the ``estimation sample'') is used to estimate treatment effects in each leaf.  The tree is pruned using cross-validation, just as in standard regression trees, but where the criterion for evaluating the performance of the tree in held-out data is based on treatment effect heterogeneity rather than predictive accuracy.

Some advantages of the causal tree method are similar to advantages of regression trees.  They are easy to explain; in the case of a randomized experiment, the estimate in each leaf is simply the sample average treatment effect.  A disadvantage is that the tree structure is somewhat arbitrary; there may be many partitions of the data that exhibit treatment effect heterogeneity, and taking a slightly different subsample of the data might lead to a different estimated partition.  The approach of estimating simple models in the leaves of shallow trees can be applied to other types of models; see  \citep{zeileis2008model} for an early version of this idea, although that paper did not provide theoretical guarantees or confidence intervals.


For some purposes, it is desirable to have a smooth estimate of $\tau(x)$.  For example, if a treatment decision must be made for a particular individual with covariates $x$, a regression tree may give a biased estimate for that individual given that the individual may not be in the center of the leaf, and that the leaf may contain other units that are distant in covariate space.  In the traditional econometrics literature, non-parametric estimation could be accomplished through kernel estimation or matching techniques. However,  their theoretical and practical properties are poor  
with many covariates. \citet{wager2017estimation} introduces {\it causal forests.} Essentially, a causal forest
is the average of a large number  of causal trees, where trees differ from one another due to subsampling.  Similar to prediction forests,
a causal forest can be thought of as a version of a nearest neighbor matching method, but one where 
there is a data-driven approach to determine which dimensions of the covariate space are important to match on.  The paper establishes asymptotic normality of the estimator (so long as tree estimation is ``honest,'' making use of sample splitting for each tree) and provides an estimator for the variance of estimates so that confidence intervals can be constructed. 

A challenge with forests is that it is difficult to describe the output, since the estimated CATE function $\hat\tau(x)$ may be quite complex.  However, in some cases one might wish to test simpler hypotheses, such as the hypothesis that the top $10\%$ of individuals ranked by their CATE have a different average CATE than the rest of the population.  \citet{chernozhukov2018generic} provides methods for testing this type of hypothesis.

As described above in our presentation of regression forests, \citet{athey2016generalized} extended the framework of causal forests to analyze nonparametric parameter heterogeneity in models where the parameter of interest can be estimated by maximum likelihood or GMM.  As an application, the paper highlights the case of instrumental variables. \citet{friedberg2018local} extends local linear regression forests to the problem of heterogeneous treatment effects, so that regularity in the function $\tau(x)$ can be better exploited.  

An alternative approach to estimating parameter heterogeneity in instrumental variables models was proposed by \citet{Hartford2016}, who use an approach based on neural nets, though distributional theory is not available for that estimator.  
Other possible approaches to estimating conditional average treatment effects can be used when the structure of the heterogeneity is assumed to take
a simple form.  Targeted maximum likelihood \citep{van2006targeted} is one approach to this, while \citet{imai2013estimating} proposed using LASSO to uncover heterogeneous treatment effects. \citet{kunzel2017meta} proposes an ML approach using ``meta-learners.''  Another popular alternative that takes a Bayesian approach is Bayesian Additive Regression Trees (BART), developed by \citet{chipman2010bart} and applied to causal inference by \citet{hill2011bayesian, green2012modeling}.

A main motivation for understanding treatment effect heterogeneity is that the CATE can be used to define policy assignment functions, that is, functions that map from the observable covariates of individuals to policy assignments.  A simple way to define a policy is to estimate $\hat\tau(x)$ and to assign the treatment to all individuals with positive values of $\hat\tau(x)$, where the estimate should be augmented with any costs of being in the treatment or control group.  \citet{hirano2009asymptotics} shows that this is optimal under some conditions.  A concern with this approach, depending on the method used to estimate $\hat\tau(x)$, is that the policy may be very complex and is not guaranteed to be smooth.  

\citet{kitagawa2015should} focus on estimating the optimal policy from a class of potential policies of limited complexity in an observational study with known propensity scores.  The goal is to select a policy function to minimize the loss from failing to use the (infeasible) ideal policy, referred to as the ``regret'' of the policy.  \citet{athey2017efficient} also studies policies with limited complexity and accomodates other constraints, such as budget constraints on the treatment, and proposes an algorthm for estimating optimal policies.  The paper provides bounds on the performance of its algorithm for the case where the data come from an observational study under confoundedness and the propensity score is unknown. The paper also extends the analysis to settings that do not satisfy unconfoundedness, for example, to settings where there is an instrumental variable. 

  \citet{athey2017efficient} shows how bringing
in insights from semi-parametric efficiency theory enables tighter bounds on performance than the ML literature, thus narrowing down substantially the set of algorithms that might achieve the regret bound.
For the case of unconfoundedness, the policy estimation procedure recommended by \citet{athey2017efficient} can be written as follows, where $\Pi$ is the set of functions $\pi:\mathbb{X}\rightarrow{0,1}$, and $\hat{\Psi_i}$ is defined above and makes use of cross-fitting as well as orthogonalization:
\begin{equation}
\label{eq:policy}
\text{max}_{\pi \in \Pi} \sum_i (2\pi(X_i) - 1) \cdot \hat{\Psi}_i 
\end{equation}
The topic of optimal policy estimation has
received some attention in the ML literature, focusing on data from observational studies with unconfoundedness, including \cite{strehl-offline, dudik-offline-1, li-offline-1, dudik-offline-2, li-offline-2, swaminathan-offline, jiang-offline, thomas-offline, kallus-offline}.  \citet{zhou2018offline} analyzes the case with more than two arms, extending the efficiency results of \citet{athey2017efficient}.

One insight that comes out of the ML approach to this problem is that the optimization problem \ref{eq:policy} can be reframed as a classification problem and thus solved with off-the-shelf classification tools.  See \citet{athey2017efficient} for details.

\section{Experimental Design, Reinforcement Learning,  and Multi-Armed Bandits}
\label{section:bandits}

ML methods have recently made substantial contributions to experimental design, with multi-armed bandits becoming more popular especially in online experiments. Thompson sampling (\citet{scott2010modern, thompson1933likelihood}) and
Upper Confidence Bounds (UCB, \citet{lai1985asymptotically}) can be viewed as a simple example of reinforcement learning (\citet{sutton1998reinforcement}) where successful assignment  decisions are rewarded by sending more units to the corresponding treatment arm.

\subsection{A/B Testing versus Multi-Armed Bandits}

Traditionally much experimentation is done by assigning a predetermined number of units to each of a number of treatment arms. Often there would be just two treatment arms. After the outcomes are measured the average effect of the treatment would be estimated using the difference in average outcomes by treatment arm. 
This is a potentially very inefficient way of experimentation where we waste units by assigning them to treatment arms that we already know with a high degreee of uncertainty to be inferior to some of the other arms. 
Modern methods for experimentation focus on balancing {\it exploration} of new treatments with {\it exploition} of treatments currently assessed to be of high quality.
Suppose what we are interested in is primarly finding a treatment that is good among the set of treatments considered, rather than in estimation of expected outcomes for the full set of treatments. Moreover, suppose that we measure the outcomes quickly after the treatments have been assigned, and suppose the units arrive sequentially for assignment to a treatment.
 After outcomes for  half the units have been observed, we may have a pretty good idea which of the treatments are still candidates for the optimal treatment. Exposing more units to  treatments that are no longer competitive is suboptimal for both exploration and exploitation purposes: it does not help us distinguish between the remaining candidate optimal treatments, and it exposes those units to inferior treatments.

Multi-armed bandit approaches 
(\citet{scott2010modern, thompson1933likelihood}) attempt to improve over this static design. In the extreme case, the assignment for each unit depends on all the information learned up to that point. Given that information, and given a parametric model for the outcomes for each treatment, and a prior for the parameters of these models, we can estimate the probability of each treatment being the optimal one. Thompson sampling suggests assigning the next unit to each treatment with probability equal to the probability that that particular treatment is the optimal one. This means that the probability of assignment to a treatment arm for which we are confident that it is inferior to some of the other treatments is low, and eventually all new units will be assigned to the optimal treatment with probability close to one.

To provide some more intuition, consider a case with $K$ treatments where the outcome is binary, so the model is a binomial distribution with treatment-arm-specific success probability $p_k$, $k=1,\ldots,K$. If the prior distribution for all probabilities is uniform, the posterior distribution for the success probability of arm $k$, given $M_k$ succeses in the $N_k$ trials concluded sofar, is Beta with parameters $M_k+1$ and $N_k-M_k+1$. Given that the Beta distribution is simple to approximate by simulation the probability that treatment arm $k$ is the optimal one (the one with the highest success probability), ${\rm pr}(p_k=\max_{m=1}^{K} p_m)$. 

We can simplify the calculations by updating the assignment probabilities only after seeing a number of new observations.
That is, we re-evaluate the assignment probabilities after a {\it batch} of new observations has come in, all based on the same assignment probabilities. From this perspective we  can view a standard A/B experiment as one where the batch is the full set of observations. From that perspective it is clear that to, at least occasionally, update the assignment probabilities to avoid sending units to inferior treatments, is a superior strategy.

An alternative approach is to use the
Upper Confidence Bounds (UCB, \citet{lai1985asymptotically}) approach. In that case we construct a 100(1-p)\% confidence interval for the population average outcome $\mu_k$ for each treatment arm. We then collect the upper bounds of these confidence intervals for each treatment arm and assign the next unit to the treatment arm with the highest value for the upper confidence bounds. As we get more and more data, we let one minus the level of the confidence intervals $p$ go to zero slowly. With UCB methods we need to be more careful with if we wish to update assignments only after batches of units have come in. If two treatment arms have very similar UCBs, assigning a large number of units to the one that has a slightly higher upper confidence bound may not be satisfactory: here Thompson sampling would assign similar numbers of units to both those treatment arms. More generally, the stochastic nature of the assignment under Thompson sampling, compared to the deterministic assignment in the UCB approach, has conceptual advantages for the ability to do randomization inference ({\it e.g.,} \citet{athey2017econometrics}).

\subsection{Contextual Bandits}

The most important extension of multi-armed bandits is to settings where we observe features of the units that can be used in the assignment mechanism. If treatment effects are heterogeneous, and that heterogeneity is associated with observed characteristics of the units, there may be substantial gains from assigning units to different treatments based on these characteristics. See
\citet{dimakopoulou2018balanced} for detials.

A simple way to incorporate covariates would be to build a parametric  model for the expected outcomes in each treatment arm (the {\it reward function}, estimate that given the current data and infer from there the probability that a particular arm is optimal for a new unit conditional on the characteristics of that unit. This is conceptually a straightforward way to incorporate characteristics, but it has some drawbacks. 
The main concern is that such methods may implicitly rely substantially on the model being correctly specified. It may be the case that the data for one treatment arm come in with a particular distribution of the characteristics, but it gets used to predict  outcomes for units with very different characteristics. 
See \citet{bayati2016} for some discussion.  A risk is that if the algorithm estimates a simple linear model mapping characteristics to outcomes, then the algorithm may suggest a great deal of certainty about outcomes for an arm in a region of characteristic space where that treatment arm has never been observed.  This can lead the algorithm to never experiment with the arm in that region, allowing for the possibility that the algorithm never corrects its mistake and fails to learn the true optimal policy even in large samples.

As a result one should be careful in building a flexible model relating the characteristics to the outcomes.
\citet{dimakopoulou2017estimation} highlight the benefits of using random forests as a way to avoid making functional form assumptions.

Beyond this issue, a number of novel considerations that arise in contextual bandits.  Because the assignment rules as a function of the features changes as more units arrive, and tend to assign more units to a given arm in regions of the covariate space where it has performed well in the past, particular care has to be taken to eliminate biases in the estimation of the reward function. Thus, although there is formal randomization, the issues concerning robust estimation of conditional average causal effects in observational studies become relevant here.  One solution, motivated by the literature on causal inference, is to use propensity score weighting of outcome models.  \citet{dimakopoulou2018balanced} studies bounds on the performance of contextual bandits using doubly robust estimation (propensity-weighted outcome modeling), and also demonstrates on a number of real-world datasets that propensity weighting improves performance.  

Another insight is that it can be useful to make use of simple assignment rules, particularly in early stages of bandits, because complex assignment rules can lead to confounding later.  In particular, if a covariate is related to outcomes and is used in assignment, then later estimation much control for this covariate to eliminate bias.  For this reason, LASSO, which selects a sparse model, can perform better than Ridge, which places weights on more covariates, when estimating an outcome model that will be used to determine the assignment of units in subsequent batches.  Finally, flexible outcome models can be important in certain settings; random forests can be a good alternative in those cases.

\section{Matrix Completion and Recommender Systems}
\label{section:matrixcompletion}

So far the methods we have discussed are primarily for settings where we observe information on a number of units in the form of a single outcome and a set of covariates or features, what is known in the econometrics literature as a cross-section setting. There are also many interesting new methods for settings that resemble what are in the econometric literature referred to as longitudinal or panel data settings. Here we discuss a canonical version of that problem, and then consider some specific methods.

\subsection{The Netflix Problem}

The Netflix Prize Competition was set up in 2006 (\citet{bennett2007netflix}), and asked researchers to use a training data set to develop an algorithm that improved on the  Netflix algorithm for recommending movies by providing predictions for movie ratings. Researchers were given a training data set that contained movie and individual characteristics, as well as movie ratings, and were asked to predict ratings for movie/individual pairs for which they were not given the ratings. Because of the magnitude of the prize, \$1,000,000, this competition and the associated problem generated a lot of attention, and the development of new methods for this type of setting accelarated substantially as a result.
The winning solutions, and those that were competitive with the winners, had some key features. One is that they relied heavily on model averaging. Second, many of the models included matrix factorization and nearest neighbor methods.

Although it may appear at first as a problem that is very distinct from the type of problem studied in econometrics, one can cast many econometric panel data in a similar form. In settings where researchers are interested in causal effects of a binary treatment, one can think of the realized data as consisting of two incomplete potential outcome matrices, one for the outcomes given the treatment, and one for the outcomes given the control treatment. Hence the problem of estimating the average treatment effects can be cast as a matrix completion problem.
Suppose we observe outcomes on $N$ units, over $T$ time periods, with the outcome for unit $i$ at time period $t$ denoted by $Y_{it}$, and a binary treatment, denoted by $W_{it}$, with
\[ \by=\left(
\begin{array}{ccccccc}
 Y_{11} & Y_{12} & Y_{13}  & \dots & Y_{1T}  \\
Y_{21}  & Y_{22} & Y_{23}   & \dots & Y_{2T}  \\
Y_{31}  & Y_{32} & Y_{33}   & \dots & Y_{3T}  \\
\vdots   &  \vdots & \vdots &\ddots &\vdots \\
Y_{N1}  & Y_{N2} & Y_{N3}   & \dots & Y_{NT}  \\
\end{array}
\right)\hskip1cm \bw=\left(
\begin{array}{ccccccc}
 1 & 1 & 0  & \dots & 1  \\
0  & 0 & 1   & \dots & 0  \\
1  & 0 & 1   & \dots & 0  \\
\vdots   &  \vdots & \vdots &\ddots &\vdots \\
1  & 0 & 1   & \dots & 0 \\
\end{array}
\right).\]
We can think of there being two matrices with potential outcomes,
\[ \by(0)=\left(
\begin{array}{ccccccc}
 ? & ? & Y_{13}  & \dots & ?  \\
Y_{21}  & Y_{22} &?   & \dots & Y_{2T}  \\
?  & Y_{32} & ?   & \dots & Y_{3T}  \\
\vdots   &  \vdots & \vdots &\ddots &\vdots \\
?  & Y_{N2} & ?   & \dots & Y_{NT}  \\
\end{array}
\right)\hskip1cm {\rm (potential\ control\ outcome)},\]
and
\[ \by(1)=\left(
\begin{array}{ccccccc}
 Y_{11} & Y_{12} & ?  & \dots & Y_{1T}  \\
?  & ? & Y_{23}   & \dots & ? \\
Y_{31}  & ? & Y_{33}   & \dots & ?  \\
\vdots   &  \vdots & \vdots &\ddots &\vdots \\
Y_{N1}  & ? & Y_{N3}   & \dots & ?  \\
\end{array}
\right)\hskip1cm {\rm (potential\ treated\ outcome)}.\]
Now the problem of estimating causal effects becomes one of imputing missing values in a matrix.

The ML literature has developed effective methods for matrix completion in settings with both $N$ and $T$ large, and a large fraction of missing data. We discuss some of these methods in the next section, as well as their relation to the econometrics literature.

\subsection{Matrix Completion Methods for Panel Data}

The matrix completion literature has focused on using low rank representations for the complete data matrix.
Let us consider the case without covariates, that is, no characteristics of the units or time periods. Let $\bl$ be the  matrix of expected values, and $\by$ the observed data matrix. The observed values are assumed to be equal to the corresponding values of the complete data matrix, possibly with error:
\[ Y_{it}=\left\{
\begin{array}{ll}
L_{it}+\varepsilon_{it} \hskip1cm &{\rm if\ } W_{it}=1,\\
0 & {\rm  otherwise.}\end{array}\right.\]
Using the singular value decomposition,
$ \bl=\bu\bs\bv^\top,$
where $\bu$ is an $N\times N$ matrix, $\bv$ is a $T\times T$ matrix, and $\bs$ is an   $N\times T$ matrix with rank $R$, with the only non-zero elements on the diagonal (the singular values). We are not interested in estimating the matrices $\bu$ and $\bv$, only in the product $\bu\bs\bv^\top$, and possibly in the singular values, the diagonal elements of $\bs$.
Obviously some regularization is required, and an effective one is to use the nuclear norm $\|\cdot\|_*$, which is equal to the sum of the singular values. Building on the ML literature (\citet{candes2009exact, mazumder2010spectral}), \citet{athey2017matrix} focus on  estimating $\bl$ by minimizing
\[ \min_{\bl} \left\{\sum_{(i,t)\in\calo} (Y_{it}-L_{it})^2+\lambda \left\| \bl\right\|_*\right\},\]
where $\lambda$ is a penalty parameter chosen through cross-validation.
Using the nuclear norm here, rather than the rank of the matrix $\bl$ is important for computational reasons. Using the
 Frobenius norm, equal to the sum of the squares of the singular values would not work because it is equal to the sum of the squared values of the matrix, and thus would lead to imputing all missing values as zeros.
For the nuclear norm case there are effective algorithms that can deal with large $N$ and large $T$. See \citet{candes2009exact, mazumder2010spectral}.

\subsection{The Econometrics Literature on Panel Data and Synthetic Control Methods}

The econometrics literature has studied these problems from a number of different perspectives.
The panel data literature traditionally focused on fixed effect methods, and has generalized those to models with multiple latent factors (\citet{bai2017principal, bai2002determining, bai2003inferential}) that are essentially the same as the low rank factorizations in the ML literature. The difference is that in the econometrics literature there has been more focus on actually estimating the factors, and using normalizations that allow for their identification. Typically it is assumed that there is a fixed number of factors.

The synthetic control literature studied similar settings, but focused on the case with only missing values for a single row of the matrix $\by$. 
\citet{Abadie2010, abadie2014} propose imputing these using a weighted average of the outcomes for other units in the same period.
\citet{doudchenko} show that the \citet{abadie2014} methods can be viewed as regression the outcomes for the last row on outcomes for the other units, and using the regression estimates from that to impute the missing values, in what 
\citet{athey2017matrix} call the vertical regression. This contrasts with a horizontal regression, common in the program evaluation literature where outcomes in the last period are regressed on outcomes in earlier periods, and those estimates are used to impute the missing values.  In contrast to both the horizontal and vertical regression approaches the matrix completion approach in principle attempts to exploit both stable patterns over time and stable patterns between units in imputing the missing values, and also can deal directly with more complex missing data patterns.

\subsection{Demand Estimation in Panel Data}

A large literature in economics and marketing focuses on estimating consumer preferences using data about their choices.   A typical paper analyzes the discrete choice of a consumer who
selects a single product from a set of prespecified imperfect
substitutes, e.g. laundry detergent, personal
computers, or cars (see, e.g., \citet{ keane2013panel} for a review). The literature typically focuses on one product category
at the time, and typically models choices among a small number of products.  Often this
literature focuses on estimating cross-price elasticities, so that counterfactuals about firm mergers or price changes can be analyzed.  Although
it is common to incorporate individual-specific preferences for observable characteristics, such
as prices and other product characteristics, there are typically a small number of latent variables
in the models.  A standard set-up starts with consumer $i$'s utility for product $j$ at time $t$, where 

$$U_{ijt} = \mu_{ij} - \phi_i p_{jt} + \epsilon_{ijt},$$ 

where $\epsilon_{ijt}$ has an extreme value distribution and is independently and identically distributed across consumers, products, and time. $\mu_{ij}$ is consumer $i$'s mean utility for product $j$, $\phi_i$ is consumer $i$'s price sensitivity, and $p_{jt}$ is the price of product $j$ at time $t$.  If the consumer selects the item with highest utility, then 

$$Pr(Y_{ijt}=j) = \frac{\exp^{\mu_{ij} - \phi_i p_{jt}}}{\sum_{j'} \exp^{\mu_{ij} - \phi_i p_{jt}}}$$

From the machine learning perspective, a panel dataset with consumer choices might be studied using techniques from matrix completion, as described above.  The model would draw inferences from products that had similar purchase patterns across consumers, as well as consumers who had similar purchase patterns across products.  However, such models would typically not be well-suited to analyze the extent to which two products are substitutes, or to analyze counterfactuals. 

For example, \cite{Jacobs2014a} propose using a related latent factorization approach
in order to flexibly model consumer heterogeneity in the context of online
shopping with a large assortment of products. They use data from medium sized online retailer. They consider
3,226 products, and
  aggregate up to the category x brand level to reduce to 440 ``products.'' 
They do not model responses to price changes or substitution between similar products; instead, in the spirit of the machine learning literature, they evaluate performance in terms of predicting which new products a customer will buy.
  

In contrast to this ``off-the-shelf'' application of machine learning to product choice, a recent literature has emerged that attempts to combine machine learning methods
with insights from the economics literature on consumer choice, typically in panel
data settings.  A theme of this literature is that models that take advantage of some
of the structure of the problem will outperform models that will not.  For example, the functional form implied by the consumer choice model from economics places a lot of structure on how products within a category interact with one another.  An increase in the price of one product affects other products in a particular way, implied by the functional form.  To the extent that the
restrictions implied by the functional form are good approximations to reality, they can greatly improve the efficiency of estimation.  Incorporating the functional forms that have been established
to be effective across decades of economic research can improve performance.

On the other hand, economic models have typically failed to incorporate all of the information that is available in a panel dataset, the type of information that matrix completion methods typically exploit.  In addition, computational issues have prevented economists from studying consumer choices across multiple product categories, even though in practice data about a consumer's purchases in one category is informative about the consumer's purchases in other categories; and further, the data can also reveal which products tend to have similar purchase patterns.  Thus, the best-performing models from this new hybrid literature tend to exploit techniques from the matrix completion literature, and in particular, matrix factorization.  

To see how matrix factorization can augment a standard consumer choice model, we can write the utility of consumer $i$ for product $j$ at time $t$ as
$$U_{ijt} = \beta'_i \theta_j - \rho'_i \alpha_j p_{jt} + \epsilon_{ijt},$$ 
where $\beta_i$, $\theta_j$, $\rho_i$, and $\alpha_j$ are each vectors of latent variables.  The vector $\theta_j$, for example, can be interpreted as a vector of latent product characteristics for
product $j$, while $\beta_i$ represents consumer $i$'s latent preferences for those characteristics.
The basic functional form for choice probabilities is unchanged, except that the utilities are now functions of the latent characteristics.

Such models had not been studied in the machine learning literature until recently, in part because the functional form for choice probabilities, which is nonlinear in a large number of latent parameters, makes computation challenging.  In contrast, traditional machine learning models might treat all products as independently chosen (e.g. \citet{Gopalan2015}), making computation much easier.  \citet{ruiz2017shopper} applies state-of-the-art computational techniques from machine learning (in particular, stochastic gradient descent and variational inference) together with a number of approximations in order to make the method scalable to thousands of consumers making choices over thousands of items in dozens or hundreds of shopping trips per consumer.  \citet{ruiz2017shopper} does not make use of any data about the categories of products; it attempts to learn from the data (which incorporates substantial price variation) which products are substitutes or complements.  In contrast, \citet{athey2017counterfactual} incorporates information about product categories and imposes the assumption that consumers buy only one product per category on a given trip; the paper also introduces a nested logit structure, which allows utilities to be correlated across products within a category, thus better accounting for consumers' choices about whether to purchase a category at all.  

A closely related approach is taken in \citet{Wan2017}.  They use a latent factorization approach that incorporates 
price variation.  They model consumer choice as a three stage process: (i) Choose whether to buy the category, (ii) Choose
  which item in category, and (iii) Choose number of the item to purchase.  The paper uses customer loyalty transaction data from two different
  datasets.
In all of these approaches, using the utility maximization approach from economics makes it possible to perform traditional analyses such as analyzing the impact of price changes on consumer welfare.
A complementary approach to one based on latent product characteristics is the work
by \citet{Semenova2018}, who considers observational
high-dimensional product attributes (e.g., text descriptions and images)
rather than latent features.



\section{Text Analysis}


There is a large machine learning literature on analyzing text data.  It is beyond the scope of this
paper to fully describe this literature; \citet{gentzkow2017text} provides an excellent recent review.  Here, we provide a high-level overview.

To start, we consider a dataset consisting of $N$ documents, indexed by $i=1,..,N$.  Each document contains a set of words.  One way to represent the data is as a $N \times T$ matrix, denoted $C$,
where $T$ is the number of words in the language, where each element of the matrix is an indicator for whether word $t$ appears in document $i$. Such a representation would los information by ignoring the ordering of the words in the text. Richer reprsentations might let $T$ be the number of bigrams, where a bigram is a pair of words that appear adjacent to one another in the document, or sequences of three of more words.

There are two types of exercises we can do with this type of data.  One is unsupervised learning, and the other is supervised.  For the unsupervised case, the goal would be to find a lower-rank representation of the matrix $C$.  Given that a low-rank matrix can be well approximated by a factor structure, as discussed above, this is equivalent to finding a set of $k$ latent characteristics of documents (denoted $\beta$) and a set of latent weights on these topics, denoted $\theta$, such that the probability that word $t$ appears in document $i$ is a function of $\theta_i' \beta_j$).  This view of the problem basically turns the problem into a
matrix completion problem; we would say that a particular representation performs well if we hold out a test set of randomly selected elements of $C$, and the model predicts well those held-out elements.  All of the methods described above for matrix completion can be applied here.

One implementation of these ideas are referred to as {\it topic models}; see \citet{blei2009topic} for a review.  These models specify a particular generative model of the data.  In the model, there are a number of topics, which are latent variables.  Each topic is associated with a distribution of words.  An article is characterized by weights on each topic.  The goal of a topic model is to estimate the latent topics, the distribution over words for each topic, and the weights for each article.  A popular model that does this is known as the Latent Dirichlet Allocation  model.

More recently, more complex models of language have emerged, following the theme that, although simple machine learning models perform quite well, incorporating problem-specific structure is often helpful and is typically incorporated in state-of-the art machine learning in popular application areas.  Broadly, these are known as {\it word embedding methods.}  These attempt to capture latent
semantic structure in language; see
\citep{Mnih2007,Mnih2012,Mikolov2013efficient,Mikolov2013distributed,
Mikolov2013linguistic,Mnih2013learning,Pennington2014,Levy2014neural,
Vilnis2014,Arora2016,Barkan2016bayesian,Bamler2017}.
Consider the neural probabilistic
language model of \citet{Bengio2003,bengio2006neural}.  That model
specifies a joint probability of sequences of words, parameterized by
a vector representation of the vocabulary.  Vector
representations of words (also known as ``distributed representations'')
can incorporate ideas about word usage and meaning
\citep{Harris1954,Firth1957,Bengio2003,Mikolov2013distributed}.

Another class of models uses supervised learning methods.  These methods are used when there is a specific characteristic the researcher would like to learn from text.  Examples might include favorability of a review, political polarization of text spoken by legislators, or whether a tweet about a company is positive or negative.  Then, the outcome variable is a label that contains the characteristic of interest.  A simple supervised learning model takes the data matrix $C$, views each document $i$ as a unit of observation, and treats the columns of $C$ (each corresponding to indicators for whether a particular word is in a document) as the covariates in the regression.  Since $T$ is usually much greater than $N$, it is important to use ML methods that allow for regularization.  Sometimes, other types of dimension reduction techniques are used in advance of applying a supervised learning method (e.g. unsupervised topic modeling).

Another approach is to think of a generative model, where we think of the words in the document as a vector of outcomes, and where the characteristics of interest about the document determine the distribution of words, as in the topic model literature.  An example of this approach is the supervised topic model, where information about the observed characteristics in a training dataset are incorporated in the estimation of the generative model.  The estimated model can then be used to predict those characteristics in a test dataset of unlabelled documents.  See \citet{blei2009topic} for more details.  

\section{Conclusion} 

There is  a fast growing machine learning literature that has much to offer empirical researchers in economics. In this review we describe some of the methods we view as most useful for economists, and that we view as important to include in the core graduate econometrics sequences. Being familar with these methods will allow researchers to do more sophisticated empirical work, and to communicate more effectively with researchers in other fields.

\bibliographystyle{plainnat}
\bibliography{references}

\end{document}